%% file: main.tex
\documentclass{article}

\usepackage{arxiv}

\usepackage[utf8]{inputenc} 
\usepackage[T1]{fontenc}    
\PassOptionsToPackage{hyphens}{url}  
\usepackage{hyperref}
\usepackage[numbers]{natbib}

\usepackage{url}            
\usepackage{booktabs}       
\usepackage{amsfonts}       
\usepackage{nicefrac}       
\usepackage{microtype}      
\usepackage{lipsum}		
\usepackage{graphicx}
\usepackage{natbib}
\usepackage{doi}

\usepackage{comment}
\usepackage{xcolor}
\usepackage{pifont}
\usepackage{graphicx}    
\usepackage{subcaption}
\title{SOK: Exploring Hallucinations and Security Risks in AI-Assisted Software Development with Insights for LLM Deployment}

\date{} 

\author{ 
    Ariful Haque \\
    Department of Cyber Physical Systems\\
    Clark Atlanta University\\
    Atlanta, GA, USA \\
    \texttt{mohdariful.haque@students.cau.edu} \\
    \And
    Sunzida Siddique \\
    Department of Computer Science and Engineering\\
    Dhaka International University\\
    Dhaka, Bangladesh \\
    \texttt{sunzida15-9667@diu.edu.bd} \\
    \And
    Md. Mahfuzur Rahman \\
    Silicon Orchard Research and Analytics Lab\\
    Dhaka, Bangladesh \\
    \texttt{mahim@siliconorchard.com} \\
    \And
    Ahmed Rafi Hasan \\
    Department of Computer Science and Engineering\\
    Dhaka, Bangladesh \\
    \texttt{rafihasan@cse.dhaka.edu} \\
    \And
    Laxmi Rani Das \\
    Department of Computer Science and Engineering\\
    Dhaka, Bangladesh \\
    \texttt{laxmi.das@cse.dhaka.edu} \\
    \And
    Marufa Kamal \\
    Department of Computer Science and Engineering\\
    Dhaka, Bangladesh \\
    \texttt{marufa.kamal@cse.dhaka.edu} \\
    \And
    Tasnim Masura \\
    Department of Cyber Physical Systems\\
    Clark Atlanta University\\
    Atlanta, GA, USA \\
    \texttt{tmasura@students.cau.edu} \\
     \And
    Kishor Datta Gupta \\
    Department of Cyber Physical Systems\\
    Clark Atlanta University\\
    Atlanta, GA, USA \\
    \texttt{kgupta@cau.edu} \\
}

\renewcommand{\shorttitle}{\textit{arXiv} Template}

\begin{document}
\maketitle

\thispagestyle{empty}

\begin{abstract}
The integration of Large Language Models (LLMs) such as GitHub Copilot, ChatGPT, Cursor AI, and Codeium AI into software development has revolutionized the coding landscape, offering significant productivity gains, automation, and enhanced debugging capabilities. These tools have proven invaluable for generating code snippets, refactoring existing code, and providing real-time support to developers. However, their widespread adoption also presents notable challenges, particularly in terms of security vulnerabilities, code quality, and ethical concerns. This paper provides a comprehensive analysis of the benefits and risks associated with AI-powered coding tools, drawing on user feedback, security analyses, and practical use cases. We explore the potential for these tools to replicate insecure coding practices, introduce biases, and generate incorrect or non-sensical code (hallucinations). In addition, we discuss the risks of data leaks, intellectual property violations and the need for robust security measures to mitigate these threats. By comparing the features and performance of these tools, we aim to guide developers in making informed decisions about their use, ensuring that the benefits of AI-assisted coding are maximized while minimizing associated risks.

\end{abstract}

\section{Introduction}

Large language models have transformed coding by enabling automatic code generation from natural language descriptions. Tools such as ChatGPT, Codium, Copilot, and Cursor AI assist developers in writing, completing, refactoring, and optimizing code to enhance productivity and improve performance ~\cite{zhang2024llm}. However, LLMs also pose security risks, as they are trained on publicly available code, which may contain insecure practices ~\cite{mohsin2024can}. This can lead to vulnerabilities in the generated code that malicious actors could exploit ~\cite{mousavi2024investigation}.
Furthermore, supply chain attacks targeting third-party services, particularly popular libraries such as npm packages, have increased dramatically in recent years ~\cite{zahan2022weak}. The use of advanced AI models in software development may unintentionally support these attacks by generating phishing messages and attack plans ~\cite{chowdhury2023chatgpt}. Despite their advantages, LLMs can also perpetuate biases and introduce bugs, which can affect performance and raise ethical concerns ~\cite{sharma2024llms}, ~\cite{dinh2024large}. LLMs offer advantages in various fields, including healthcare, education, programming, improving communication, workflows, and knowledge discovery ~\cite{thapaleveraging}, ~\cite{zernikow2023clinical}. Although LLMs cannot replace human programmers, they streamline tasks such as code writing, documentation, and bug detection, making them increasingly valuable in programming ~\cite{sadik2023analysis}. However, the widespread use of LLMs raises concerns about possible security vulnerabilities and hallucinations ~\cite{jungherr2023using}. This paper explores these trade-offs with the guidance of developers in leveraging AI’s benefits while mitigating associated risks.
\subsection{Related work}

Heibel et al. ~\cite{heibel2024mapping} discuss Malicious Programming Prompt (MaPP) attacks, where attackers manipulate prompts to make large language models (LLMs) generate vulnerable code. Despite improvements in model capabilities, these attacks remain effective across various platforms. The authors tested MaPP on seven LLMs using HumanEval and Common Weakness Enumerations (CWEs) and found that simple prompts induced vulnerabilities without affecting code correctness. In addition, the best-performing models were more prone to malicious instructions. Their findings suggest that enhancing models alone is not enough to prevent prompt manipulation attacks.

Similarly another author Derner et al. ~\cite{derner2023beyond}  analyzes the security risks of large language models (LLMs) like ChatGPT, focusing on vulnerabilities in content filters and the potential for malicious use. This paper identifies risks such as harmful text generation, data leaks, and unethical content creation. It also highlights the importance of informing policymakers and industry professionals about these issues. The authors evaluate the effectiveness of ChatGPT’s content filters and explore the ethical concerns surrounding its use. They conclude that current safeguards are insufficient, allowing for harmful outputs. The study recommends further research to strengthen safeguards and investigate the broader societal effects of LLMs.

In another paper, Oviedo et al. ~\cite{oviedo2023risks} discuss the advanced AI language model ChatGPT, which is widely used for tasks like customer service and chatbots. However, it has drawbacks, including the potential to provide incorrect or unsafe information that may impact user safety. This paper also evaluates ChatGPT’s overall effectiveness in promoting safety in contexts such as phone usage while driving and stress management in the workplace.

Alternatively, Jacobi et al. ~\cite{jacobi2024ai} emphasize that the increasing complexity and frequency of cyber threats necessitate updated approaches to strengthen Governance, Risk, and Compliance (GRC) frameworks. One promising approach involves the use of artificial intelligence, particularly LLMs like ChatGPT and Google Gemini, to enhance cybersecurity guidance. Research indicates that ChatGPT generally provides more relevant, accurate, and context-appropriate advice compared to Google Gemini. While both models have limitations, this study highlights the potential benefits of integrating LLMs into GRC frameworks, especially when combined with human expertise to address complex issues.

In another case, Atzori et al. ~\cite{atzori2024evaluating} examine on LLMs such as ChatGPT, GPT-4, Claude, and Bard can be misused to create phishing attacks. These models can generate realistic phishing emails and websites without modification, allowing attackers to easily scale their efforts using malicious prompts. To counter this threat, researchers have developed a BERT-based detection tool that effectively identifies phishing prompts across various LLMs. BERT-based detection tool that effectively identifies phishing prompts across various LLMs.

Another author, Latif et al. ~\cite{siddiq2023generate} discuss previous research in code generation using Large Language Models (LLMs), which usually focus on functional correctness but overlook security. It introduces the SALLM framework, which adds security-specific prompts and metrics to evaluate secure code generation. While earlier studies mainly checked syntactic and functional accuracy. This paper improves on that by using a rule-based repair system to enhance syntactic correctness, resulting in better compilation rates, especially for GPT-4. However, the study acknowledges challenges like potential bias in manually created prompts and the lack of real-world task representation, suggesting that future research should improve datasets and security evaluation methods.

Mishra et al. ~\cite{mishra2024fine}introduced FAVA-Bench dataset for fine-grained hallucination detection in language models (LMs), featuring a comprehensive taxonomy of hallucination types. The dataset includes span-level human annotations and responses from models like Llama2-Chat and ChatGPT. It comes from a number of different sources, such as the No Robots dataset, the WebNLG dataset, and Open Assistant queries. They developed FAVA, a retrieval-augmented LM fine-tuned for detecting and mitigating hallucinations, which significantly outperformed existing systems like GPT-4. The study highlighted that over 60\% of errors in model outputs were unverifiable, emphasizing the prevalence of hallucinations. While the work provides a robust framework, its focus on specific models limits generalizability, prompting future research to adapt FAVA for broader fact-checking tasks and evaluate it across diverse LM architectures.

\subsection{Our Contribution}
Key Contributions of our SoK paper are:
\textbf{User-Centric Insights:} Incorporates feedback from IT professionals to evaluate the impact of AI tools on productivity, error reduction, and collaboration while addressing issues like hallucinations and contextual errors. (in Sections 2-3).

\textbf{Evaluation of AI Coding Tools:} Provides a comprehensive analysis of tools like GitHub Copilot, ChatGPT, Cursor AI, and Codeium AI, detailing their capabilities, user benefits, and limitations in improving software development workflows. (in Section 3)
    
 \textbf{Security and Risk Analysis:} Identifies critical vulnerabilities, including data leaks, adversarial attacks, and replication of insecure coding practices, and proposes strategies to mitigate these risks. (in sections 4-5)
    
\section{User Feedback Data}

This dataset was collected from a survey to gather feedback on AI coding tools. We use Copilot AI, Codium AI, Cursor AI, and ChatGPT as tools. The survey aimed to understand user experience, satisfaction, and areas for improvement. The data set includes responses from 66 individuals. Respondents are collected from a renowned IT company, representing various departments and teams, including the Machine Learning/AI Team, Web Development Team, Mobile Development Team, Software Quality Assurance (QA) Team, Human Resources (HR) \& Administration, and Marketing. In our analysis, we evaluate the tool features based on feedback to highlight their strengths and areas for improvement ~\ref{tab:ai_tool_comparison}.

\begin{table*}[!h]
\centering
\begin{tabular}{p{4cm}cccc}
\toprule
\textbf{Features}         & \textbf{Cursor AI Rating} & \textbf{Codeium AI Rating} & \textbf{ChatGPT Rating} & \textbf{Copilot Rating} \\ \midrule
\textbf{Code Generation}   & 3.70                      & 3.24                       & 4.03                    &         4.14          \\
\textbf{Code Refactoring}  & 3.59                      & 3.30                       & 3.90                    &     4.0               \\
\textbf{Code Debugging}    & 3.66                      & 3.19                       & 3.90                    &      4.0             \\
\textbf{Code Explanation}  & 3.77                      & 3.30                       & 4.20                    &      4.14             \\
\textbf{Code AutoComplete} & 3.88                      & 3.38                       & \textit{N/A}            &    4.29              \\ \bottomrule
\end{tabular}
\caption{Comparison of AI Tool Ratings Across Key Features}
\label{tab:ai_tool_comparison}
\end{table*}

Analysis shows \textbf{ChatGPT} excelled in features of \emph{Code Generation}, \emph{Code Refactoring}, and \emph{Code Explanation}, making it a versatile and highly capable tool for developers seeking comprehensive support. \textbf{Cursor AI} demonstrated balanced performance across all categories, leading specifically in \emph{Code AutoComplete} with a high rating of \textbf{3.88}. \textbf{Codeium AI}, while consistent, scored the lowest overall, indicating areas where improvements are needed to compete effectively with other tools. \textbf{Copilot} delivered strong results across the board, particularly in \emph{Code Explanation} (\textbf{ 4.00}) and \emph{Code AutoComplete} (\textbf{3.92}), demonstrating its ability as a multifaceted developer assistance provider. Below is a summary of the key insights and feedback from user responses regarding their experiences with various AI coding tools.

\begin{figure}[h!]
    \centering
    \includegraphics[width=0.6\textwidth]{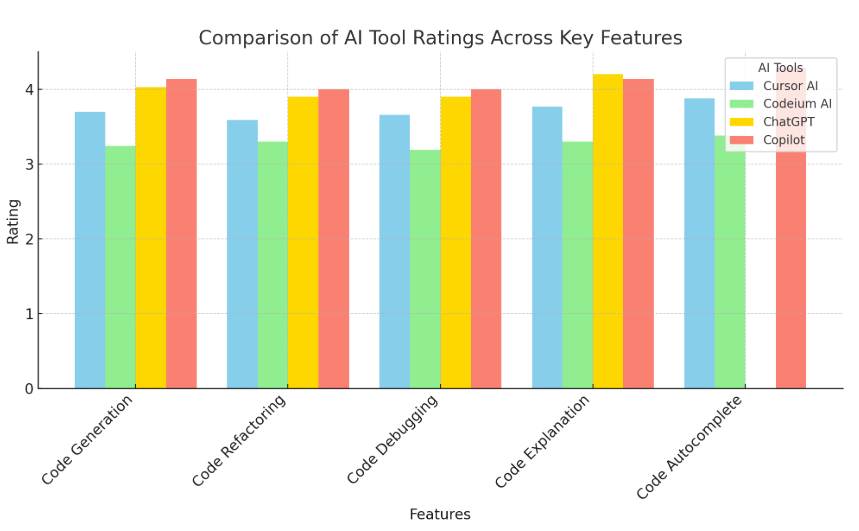}  
    \caption{Comparing the ratings of different AI tools}
    \label{fig:compare}
\end{figure}

We have analyzed the feedback for each tool to determine which tool provided the best responses. The sentiment counts for each ChatGPT, Codeium, Copilot, and Cursor AI model are categorized as positive, negative, or neutral. Below is the sentiment distribution based on user feedback:
\begin{figure}[h!]
    \centering
    \includegraphics[width=0.6\textwidth]{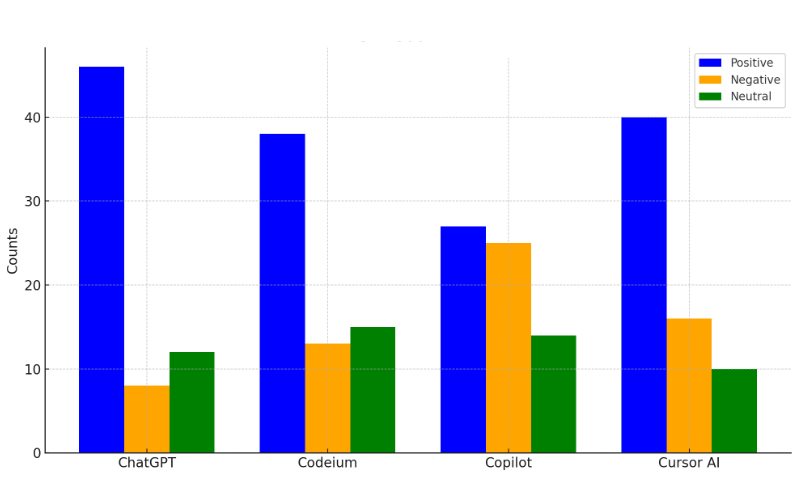}  
    \caption{Each satisfaction Sentiment Counts}
    \label{fig:poseneg}
\end{figure}

\begin{figure}[h!]
    \centering
    \includegraphics[width=0.4\textwidth]{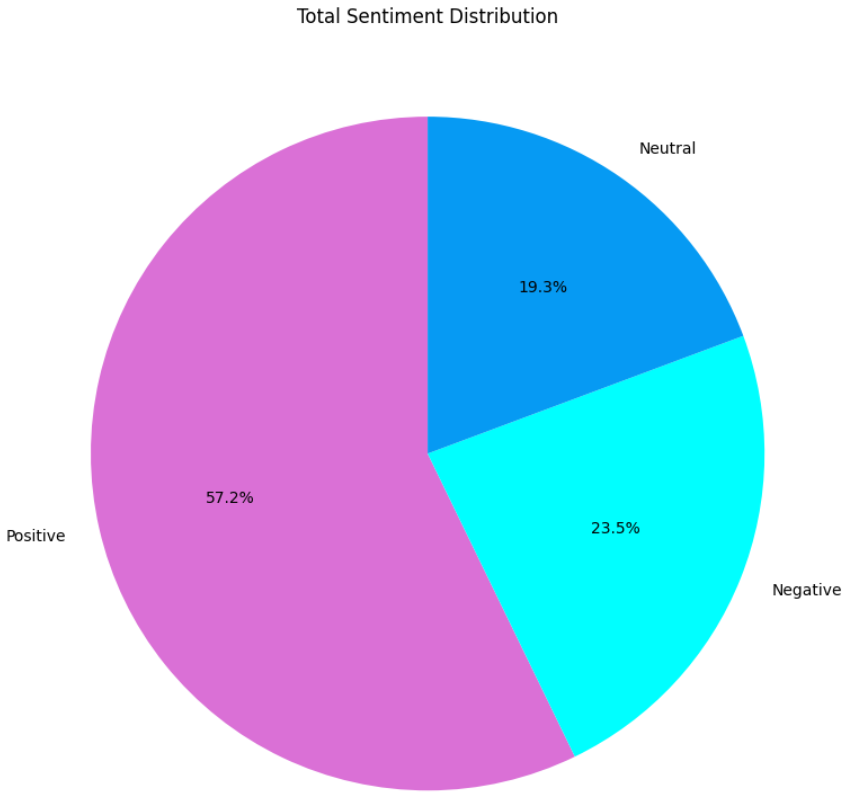}  
    \caption{Each satisfaction Sentiment Counts}
    \label{fig:over}
\end{figure}

Based on the feedback the analysis found that ChatGPT received the highest positive sentiment, with 46 positive ratings, 8 negative ratings, and 12 neutral ratings. Users provide feedback on ChatGPT for its accuracy and natural language processing capabilities, making it a powerful tool for various tasks. However, some users pointed out occasional issues with context understanding in complex scenarios. Conversely, Codeium AI followed with 38 positive responses, 13 negative responses, and 15 neutral responses. Users appreciated Codeium's coding support and speed, but some users found its responses lacking in-depth documentation. On the other hand, Copilot received a more mixed response, with 27 positive, 25 negative, and 14 neutral ratings. While it was valued for its integration with development environments and helpful suggestions, users observed that it occasionally produced irrelevant or incomplete code. Lastly, Cursor AI provides 40 positive responses, 16 negative responses, and 10 neutral responses. Its ease of use and ability to handle specific tasks such as design and content generation, although some users expressed concerns about occasional misinterpretation of input. Overall, ChatGPT received the highest ratings among AI tools, while Copilot experienced a more balanced response with a significant proportion of negative feedback. Below a figure is ~\ref{fig:over} illustrated for overall sentiment distribution, with "Positive" highlighted for better visibility.

These AI tools (ChatGPT, Codeium, Copilot, and Cursor AI) are making a significant impact by enhancing productivity, efficiency, and creativity across various domains. Analyzing sentiment data reveals a remarkable trend: the majority of feedback highlights positive experiences with these tools. This figure ~\ref{fig:over} reveals that AI tools largely contribute to user satisfaction.

\input{background}

\section{LLM  Security Issue}
LLM security is a comprehensive set of practices that are designed to protect large language models from potential threats and vulnerabilities ~\cite{kavian2024llm}. To protect this security, regular code reviews and concurrent programming play a crucial role. Additionally, we can reduce security risks by using OWASP guidelines and using trusted libraries. Furthermore, to block malicious data and enhance security, implementing user input validation and sanitization processes is very essential. Data encryption, employing methods such as AES and TLS, ensures data security and safeguards any systems from unauthorized access. Furthermore, we need to maintain security logs and monitor suspicious activity to significantly reduce the attacks. For example,

\begin{figure*}[ht!]
    \centering
    
    \begin{subfigure}[b]{0.47\textwidth}
        \centering
        \includegraphics[width=\textwidth, height=0.2\textheight]{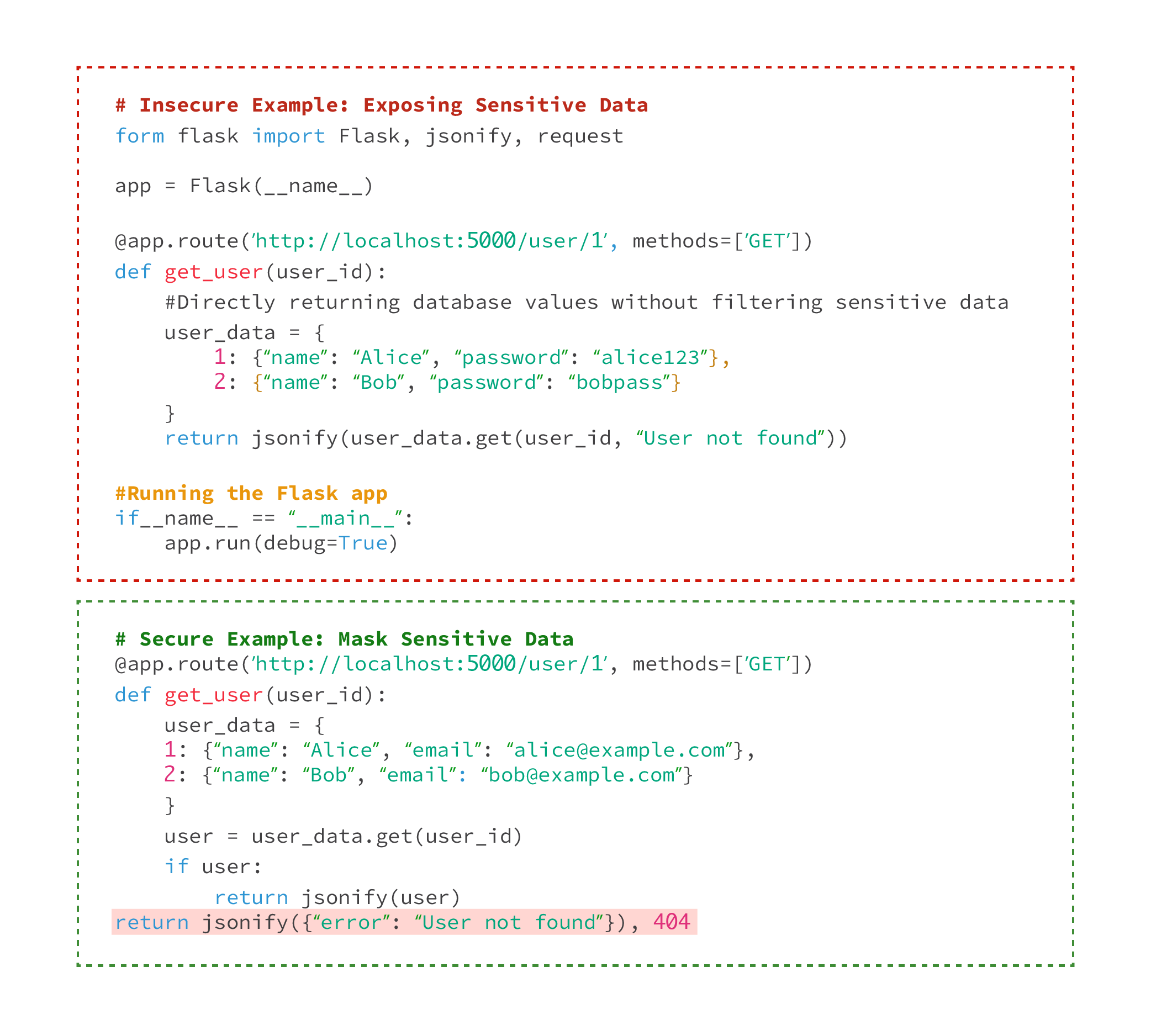}
        \caption{API Endpoint Vulnerabilities}
        \label{fig:gllm1}
    \end{subfigure}
    \hfill 
    \begin{subfigure}[b]{0.47\textwidth}
        \centering
        \includegraphics[width=\textwidth, height=0.2\textheight]{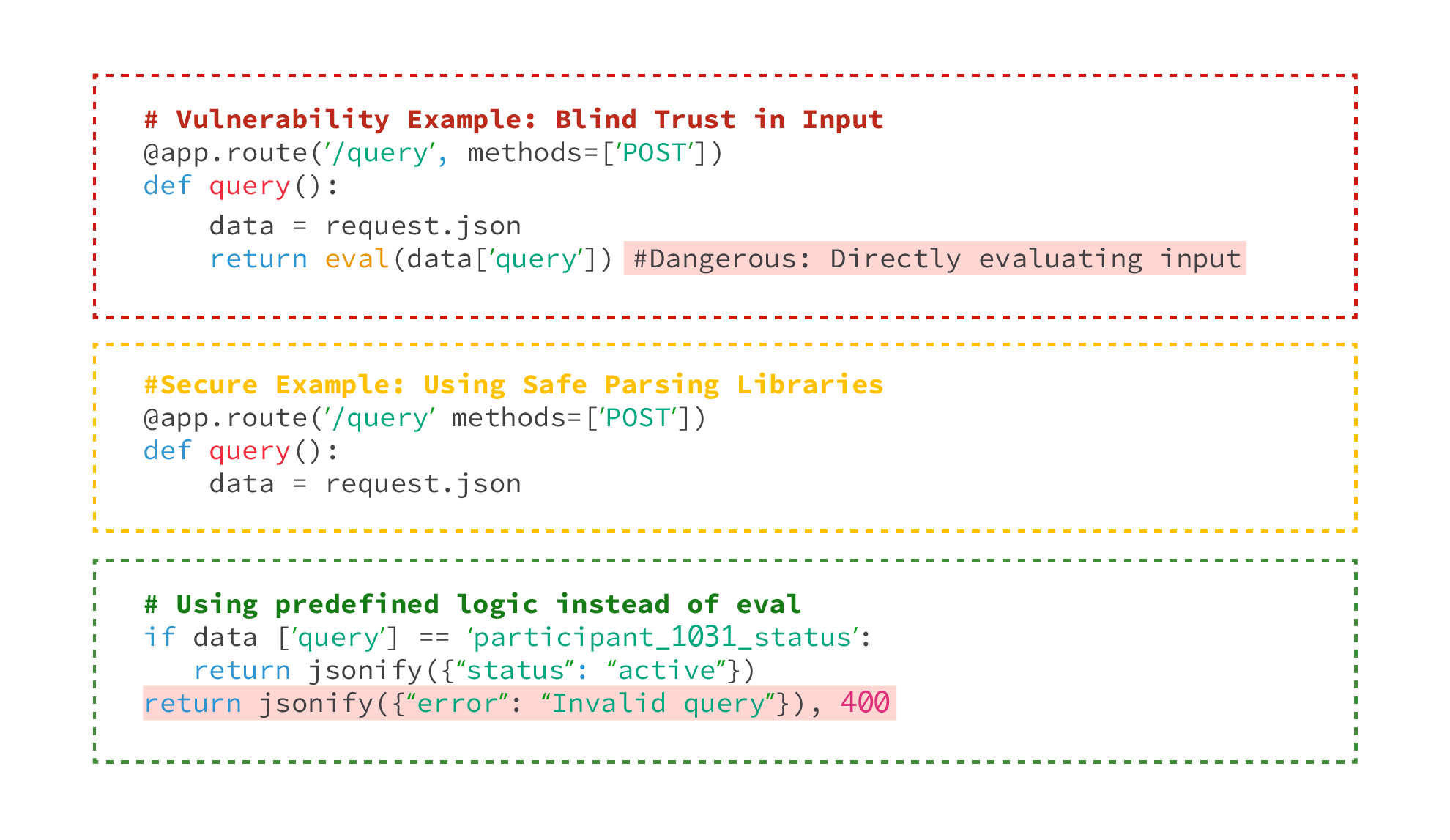}
        \caption{Transition in Queries from Participant 1031}
        \label{fig:gllm22}
    \end{subfigure}

    \caption{Examples of LLM Security}
    \label{fig:gllm}
\end{figure*}

This picture \ref{fig:gllm1} conveys the user creation API endpoint has serious vulnerabilities, as it uses proper authentication and MD5 hashing, which is dangerous for security. Additionally, figure~\ref{fig:gllm22} shows two prompts from Participant 1031 included for security rather than querying the AI assistant. Deploying firewalls, and intrusion detection systems, and providing ongoing security training for developers are essential for LLM security.

\subsection{Key Components of  LLM Security Strategy}

LLM security focuses on four main areas: data security, model security, infrastructure security, and ethical concerns ~\cite{confidentai2024llmsecurity}. Securing these areas involves a combination of standard cybersecurity practices and LLM-specific protections. Data security involves mitigating risks like data leakage, poisoning, and privacy breaches through robust measures, including encryption, access control, and protocols for ensuring data integrity. Model security focuses on challenges such as misinformation, hallucinations, and denial-of-service attacks, advocating for the adoption of authentication protocols, tamper protection, and thorough validation processes. Infrastructure security emphasizes the need for securing hosting environments with firewalls, encryption, and physical safeguards to protect against both digital and physical threats. Ethical considerations address concerns like bias, toxicity, and discrimination, highlighting the importance of ethical guidelines and responsible practices to ensure fairness and accountability. A comprehensive approach to these dimensions is essential for integrating LLMs securely, reliably, and responsibly into various applications. Data security is crucial for protecting sensitive training data, user inputs, and maintaining data integrity ~\cite{kavian2024llm}. Securing LLMs requires robust access control, encryption, and monitoring to prevent breaches and unauthorized modifications ~\cite{confidentai2024llmsecurity}. Infrastructure hosting LLMs must also be safeguarded against cyber threats to mitigate vulnerabilities ~\cite{llmsecurity}. Compliance with regulations like GDPR and HIPAA is essential to minimize legal risks and uphold organizational reputation in sensitive environments ~\cite{kavian2024llm}.

\subsection{LLM code Vulnerabilities}
LLM Code Vulnerabilities are security issues or weaknesses that appear in code produced by the Large Language Model ~\cite{llmsecurity}. Code problems can occur for a variety of reasons, including technical errors, human error, open-source software (OSS) reuse, and even unexpected zero-day attacks. Some example are discussed below:

\begin{figure*}[h]
    \centering
    
    \begin{subfigure}[b]{0.47\textwidth} 
        \centering
        \includegraphics[width=\textwidth, height=0.3\textheight]{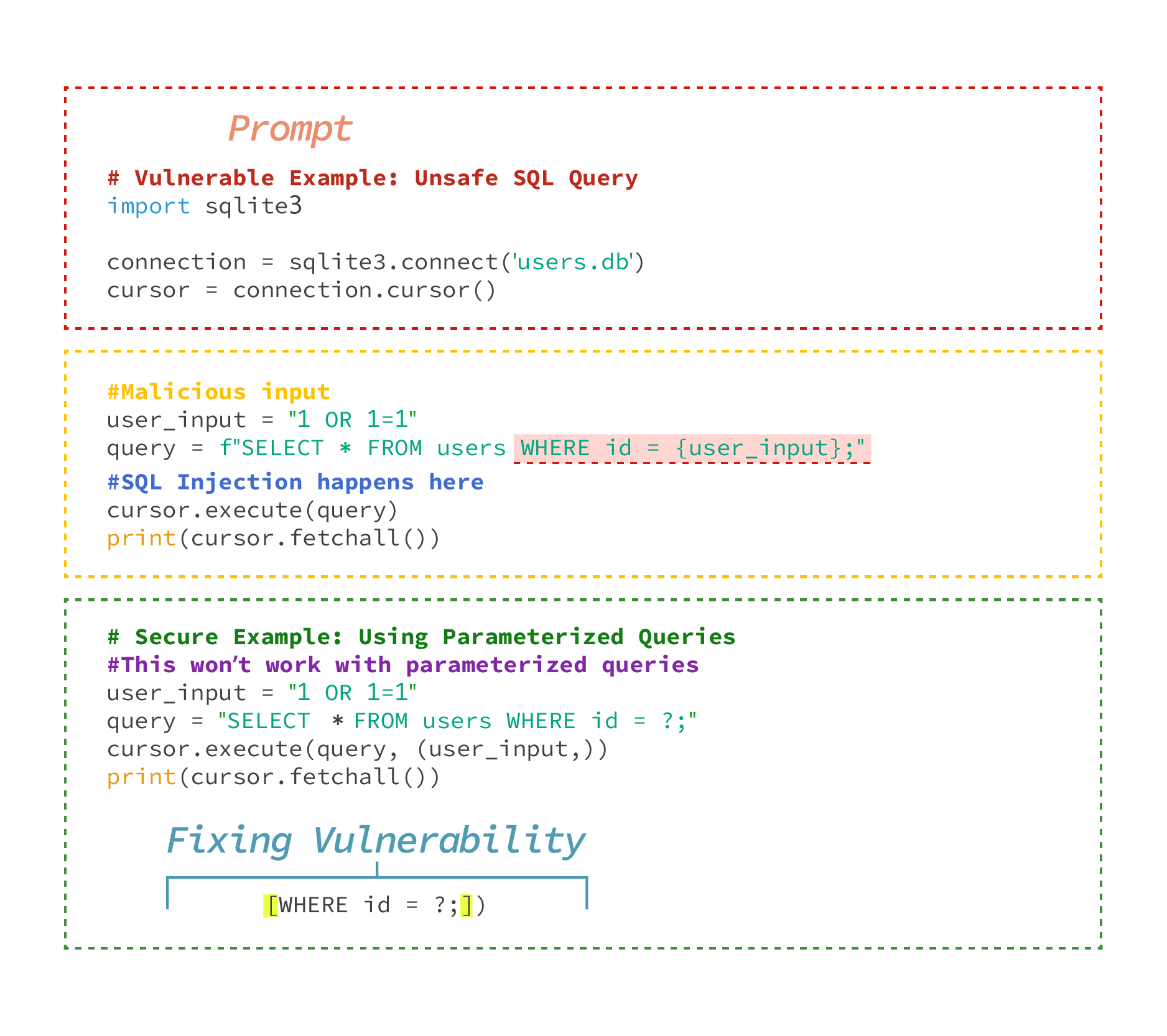} 
        \caption{SQL Injection vulnerabilities}
        \label{fig:gllm33}
    \end{subfigure}
    \hfill 
    \begin{subfigure}[b]{0.47\textwidth} 
        \centering
        \includegraphics[width=\textwidth, height=0.3\textheight]{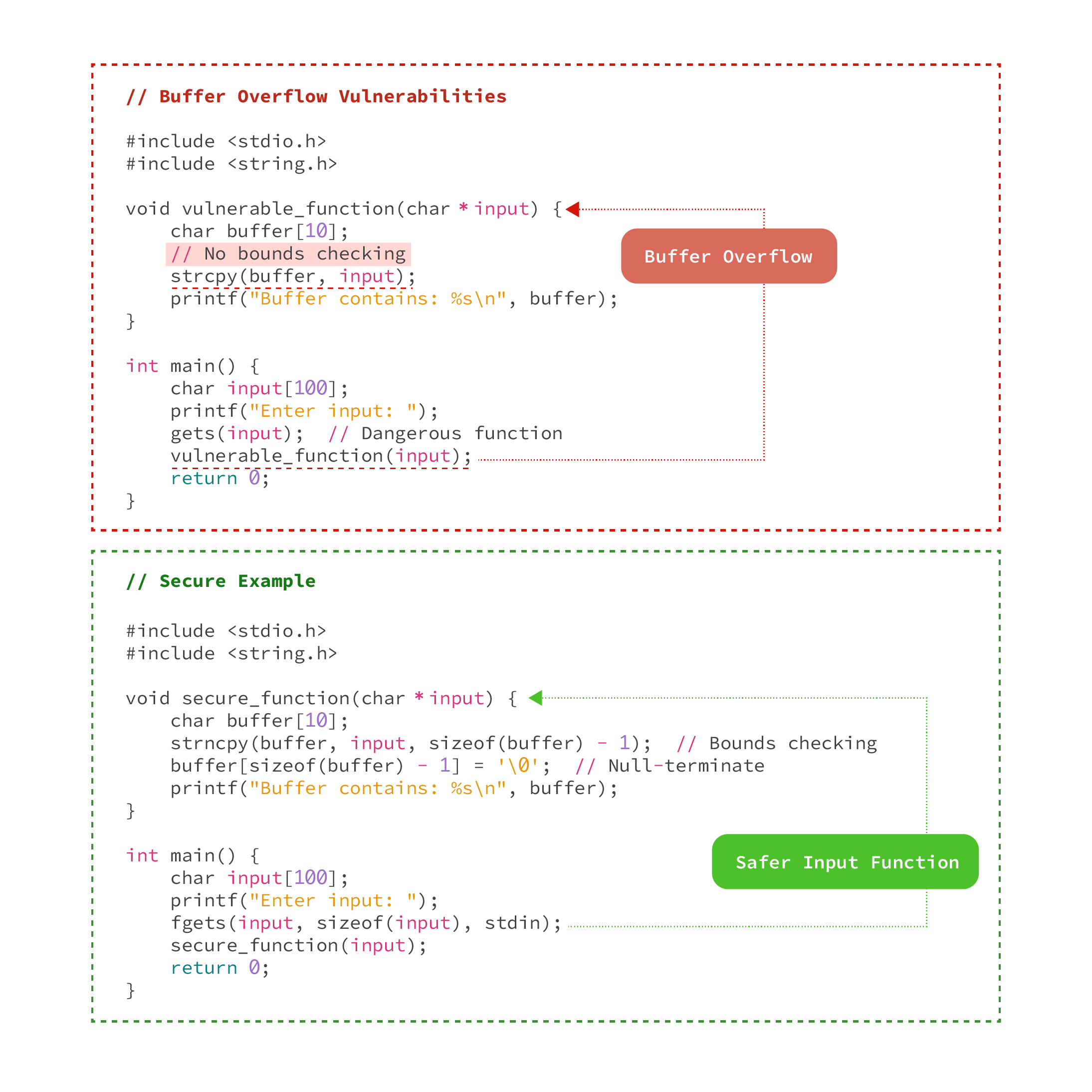} 
        \caption{Buffer Overflows vulnerabilities}
        \label{fig:gllm44}
    \end{subfigure}

    \caption{Examples of LLM Vulnerability}
    \label{fig:gllm222}
\end{figure*}

In the first image ~\ref{fig:gllm33}, this code exposes a SQL injection vulnerability by directly attaching untested data to SQL queries, specifically the username variable. This allows hackers to manipulate the input and execute SQL commands. Additionally, another code ~\ref{fig:gllm44} snippet demonstrates a buffer overflow vulnerability due to the lack of checks when copying elements from src to dest. LLM sometimes produces code that uses older programming methods and libraries ~\cite{OWASP2024}, which are not compliant with modern security. This increases the security risk. Although LLMs are designed to generate useful code, sometimes faulty code is regenerated from the training dataset, which weakens the security of the application ~\cite{llmsecurity}. On the other hand, malicious users can create malicious code by manipulating input prompts, such as making minor changes to cause errors ~\cite{confidentai2024llmsecurity}. Additionally, errors in LLM's training data also create major vulnerabilities in the code, especially when errors occur systematically ~\cite{Acorn2024}. LLM often lacks the deep contextual understanding needed to create secure code, resulting in sometimes irrelevant advice for specific security contexts ~\cite{OWASP2024}. Typically, they are not updated with the latest security vulnerabilities and threats, leaving the generated code open to new attacks ~\cite{confidentai2024llmsecurity}. Moreover, errors in generated code go undetected due to lack of security checks, which can cause serious security problems ~\cite{abdali2024securing}.

\subsection{LLM hallucination}
 LLM hallucination refers to instances when a language model, such as ChatGPT, generates information that is incorrect, irrelevant, or nonsensical ~\cite{liu2024exploring}. This phenomenon can lead to misleading or false outputs that may confuse users or propagate misinformation. A comprehensive taxonomy of hallucination types and issues in Generative Large Language Models focusing on errors in code generation ~\cite{microsoft_privacy_statement}. Hallucinations in LLMs occur when models produce incorrect, inconsistent, or nonsensical outputs, undermining functionality and reliability ~\cite{sentinelone_chatgpt_security} ~\cite{codecademy_hallucinations_ai}. Key issues include intent conflicts, where generated code misaligns with overall task goals (overall semantic conflicting) or with specific local intent (local semantic conflicting) ~\cite{wang2024your}. Context deviations manifest as logical inconsistencies, repetitive code, dead code, or context mismatches ~\cite{cursor_security} ~\cite{cursor_privacy2}. Knowledge errors arise from misuse of APIs (incorrect API calls, non-existent methods) and undefined or misused identifiers (variable misnaming, undefined references). Further, expression issues involve incorrect constants, faulty logic ~\cite{llmsecurity}in loops, conditions, or branches, and redundant copying of input contexts. Additional errors include IO/assert statement errors, such as incorrect function definitions (wrong parameters, return type mismatches) and improper assignments that disrupt logic. Problems with libraries and parameters encompass missing or unnecessary libraries, incorrect library imports, and mismatched parameters (extra arguments, missing function parameters). Identifier issues include undefined, misused, or duplicated identifiers (name collisions, wrong scope). Other critical types include semantic misalignment, where generated code produces unintended side effects, and efficiency issues, such as generating performance-inefficient code or overuse of computational resources. Additionally, output structuring errors like improper formatting, invalid indentation, or broken comments can lead to reduced code readability. Addressing these hallucination types is essential for improving LLMs by refining training datasets, enhancing contextual comprehension, and integrating robust validation systems to identify and correct hallucinations. Also, LLMs sometimes give nonsensical answers that are completely unrelated to the prompt ~\cite{liu2024exploring}. For example, if the prompt asks "Thomas Edison was born in 1847" and it answers "Edison was born in 1947" then this creates a conflict. One study found that 14.3\% of ChatGPTs had this type of problem. Another problem is that LLMs often generate context-free or random data ~\cite{chybiskov_2023_hallucinations}.

\subsection{Case study of LLM hallucinations}

\begin{figure*}[htbp]
    \centering

    \begin{subfigure}[b]{0.3\textwidth}
        \centering
        \includegraphics[width=1\textwidth, height=0.2\textheight]{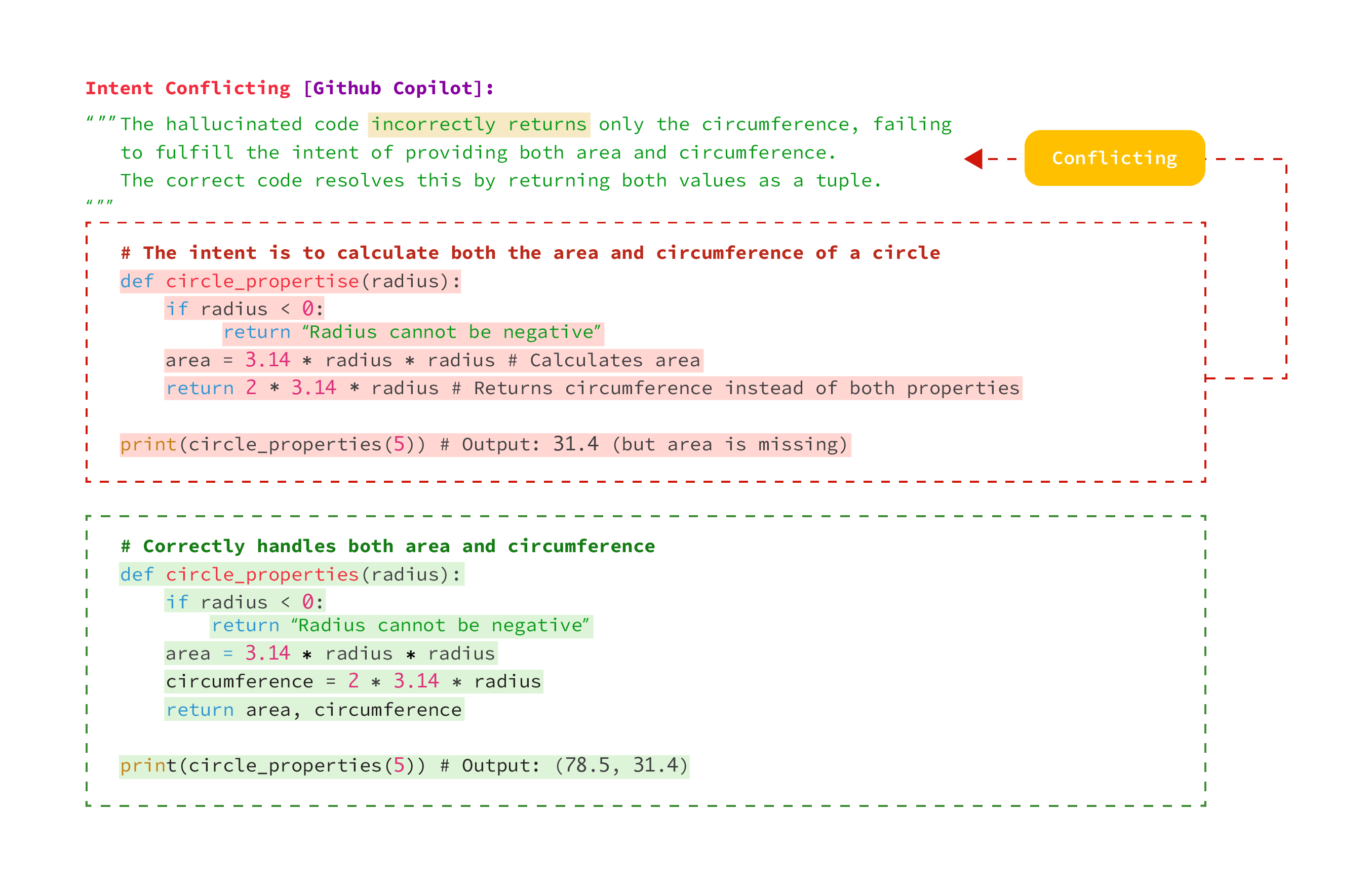} 
        \caption{Intent Conflict}
        \label{fig:gllm5}
    \end{subfigure}
    \hfill
    \begin{subfigure}[b]{0.3\textwidth}
        \centering
        \includegraphics[width=1\textwidth, height=0.2\textheight]{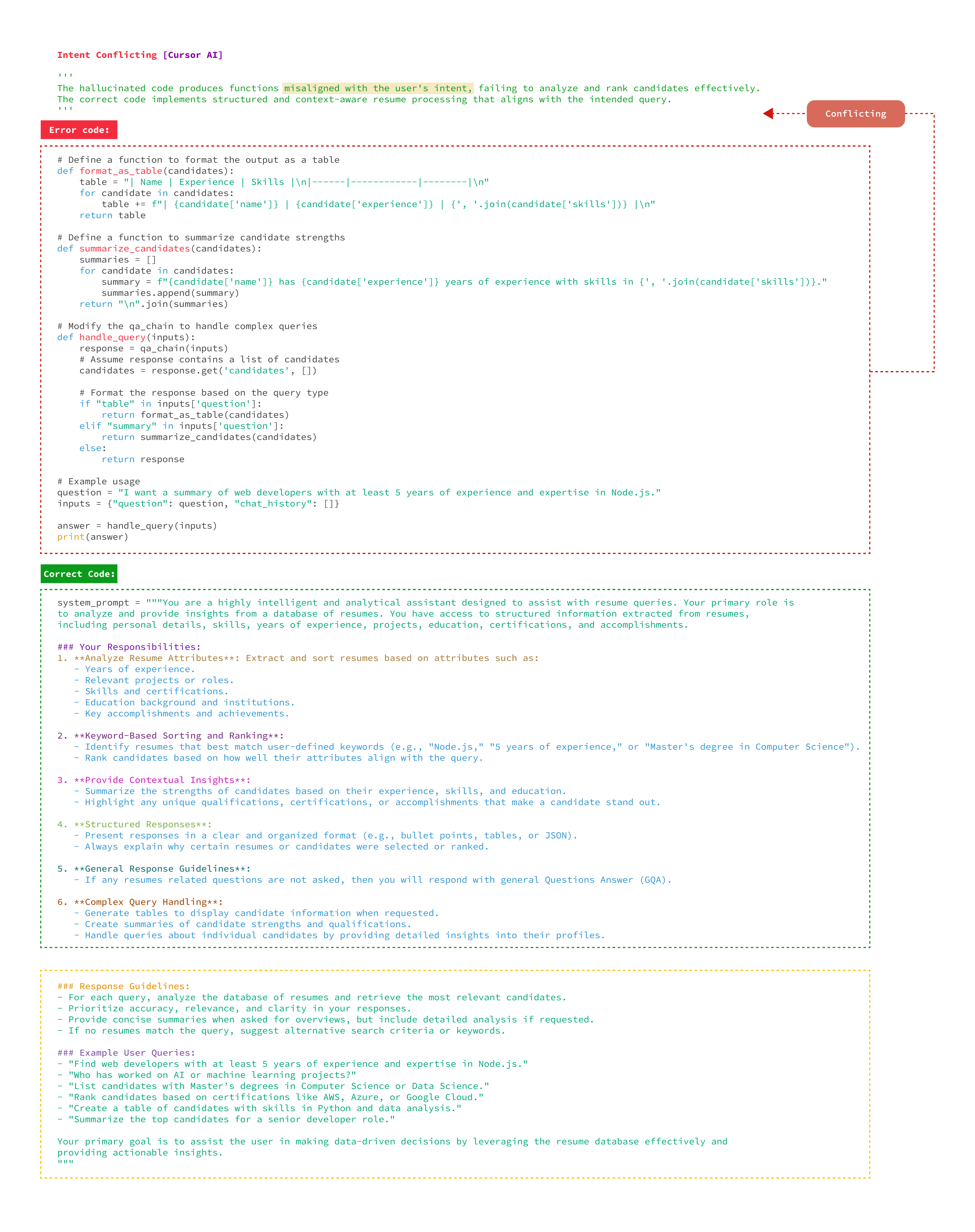} 
        \caption{semantic conflicting}
        \label{fig:gllm15}
    \end{subfigure}
    \hfill
    \begin{subfigure}[b]{0.3\textwidth}
        \centering
        \includegraphics[width=1\textwidth, height=0.2\textheight]{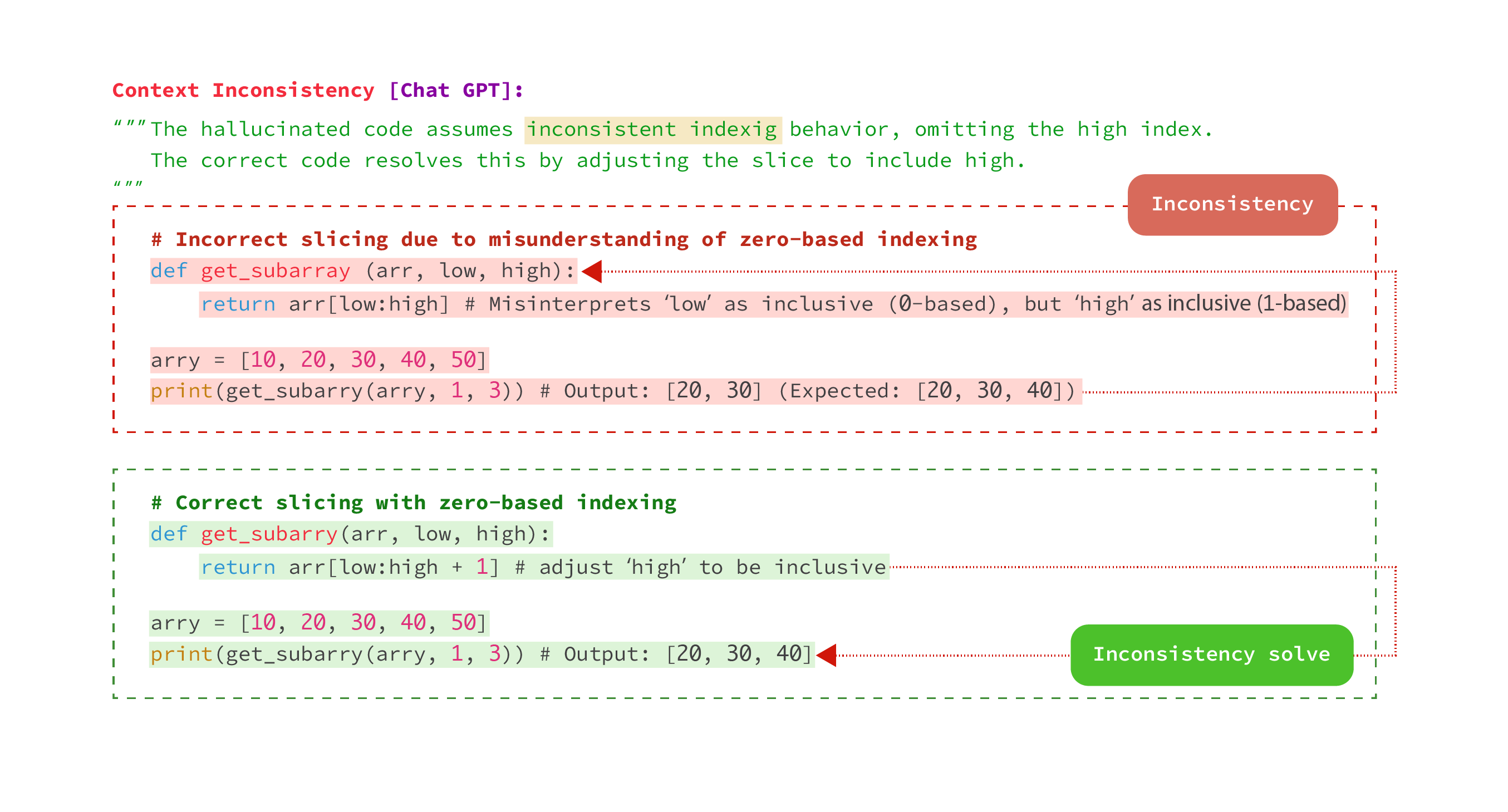} 
        \caption{Context Inconsistency}
        \label{fig:gllm6}
    \end{subfigure}
    \hfill
    \begin{subfigure}[b]{0.3\textwidth}
        \centering
        \includegraphics[width=1\textwidth, height=0.2\textheight]{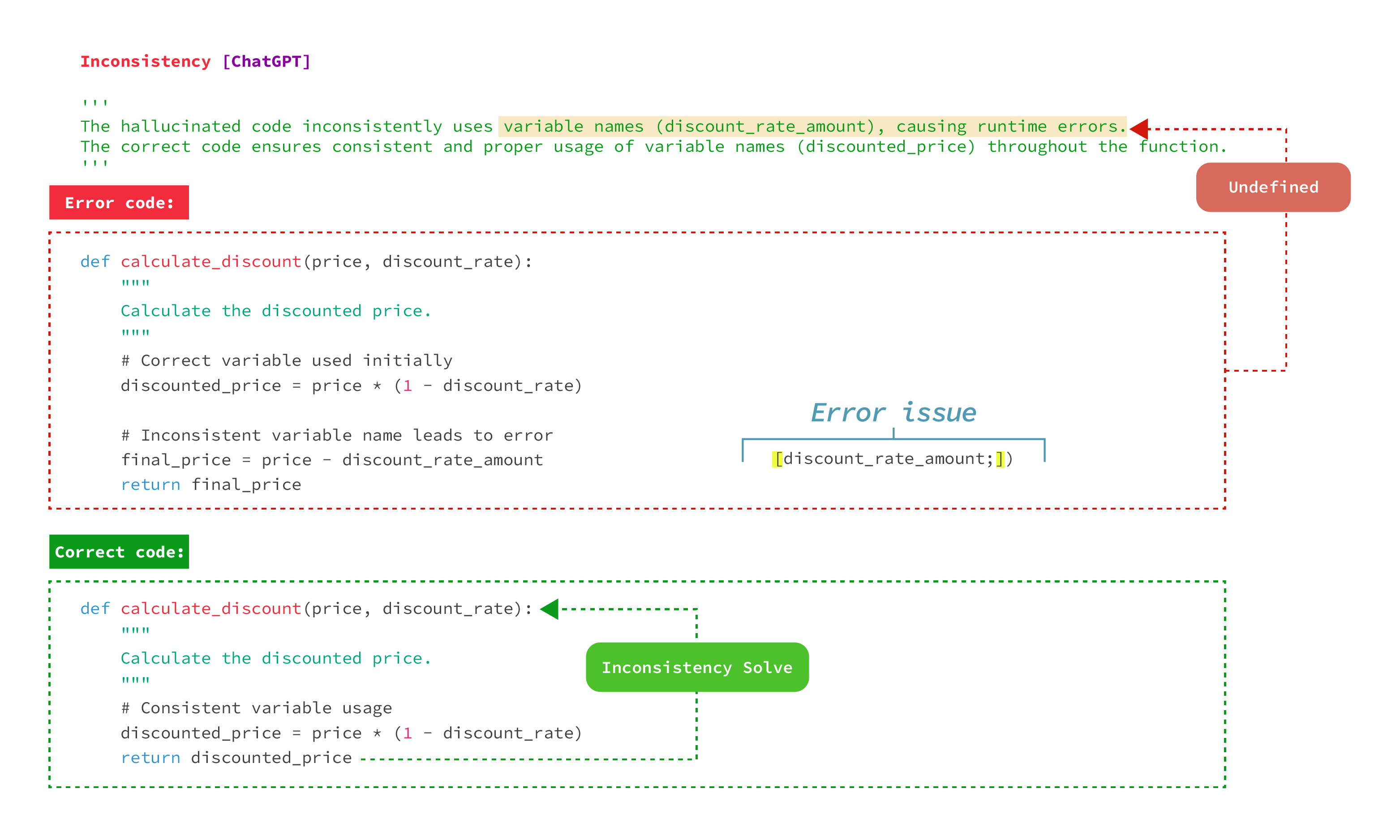} 
        \caption{Expression Inconsistency}
        \label{fig:gllm16}
    \end{subfigure}
    \hfill
    \begin{subfigure}[b]{0.3\textwidth}
        \centering
        \includegraphics[width=1\textwidth, height=0.2\textheight]{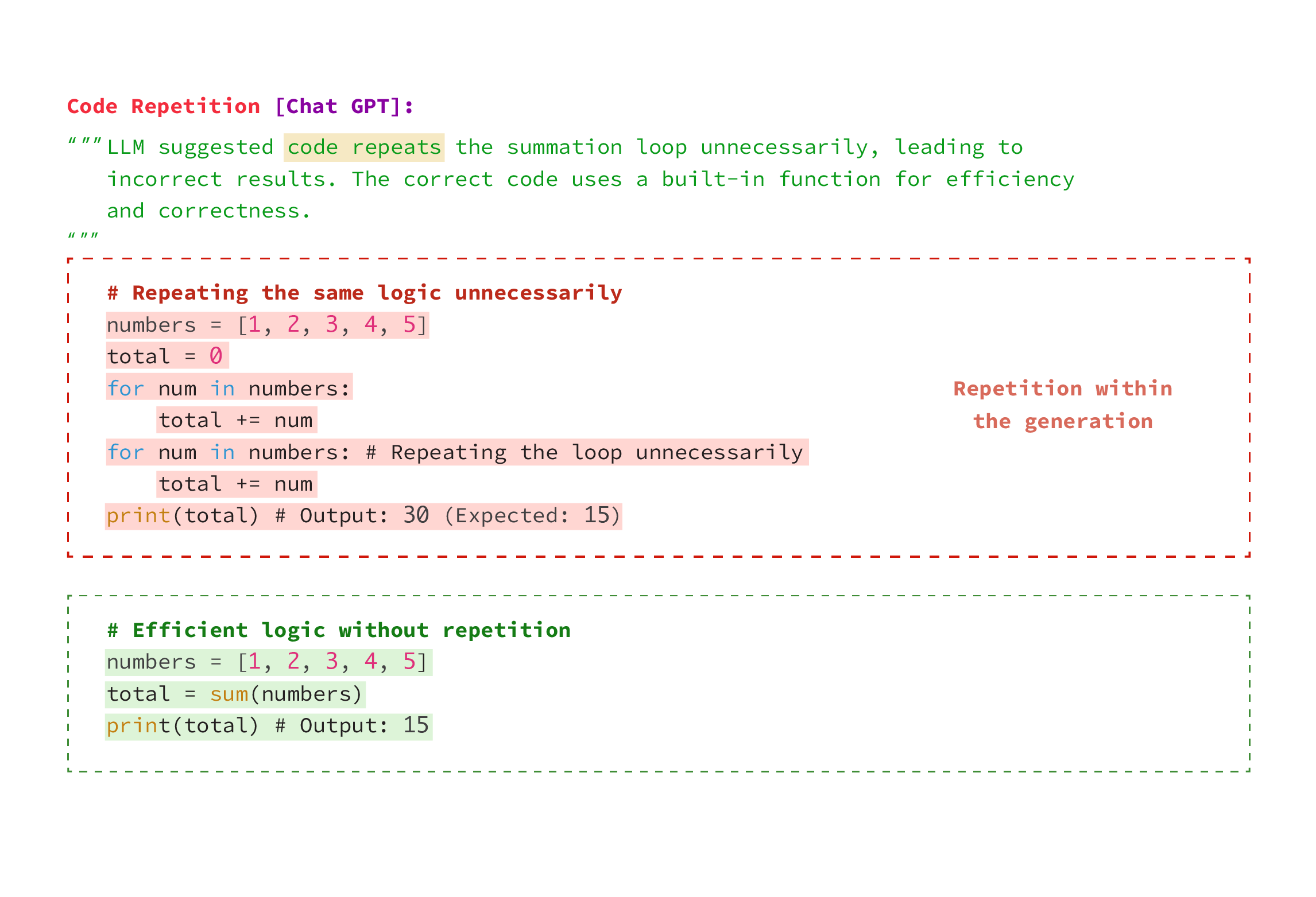} 
        \caption{Code Repetition}
        \label{fig:gllm7}
    \end{subfigure}
    \hfill
    \begin{subfigure}[b]{0.3\textwidth}
        \centering
        \includegraphics[width=1\textwidth, height=0.2\textheight]{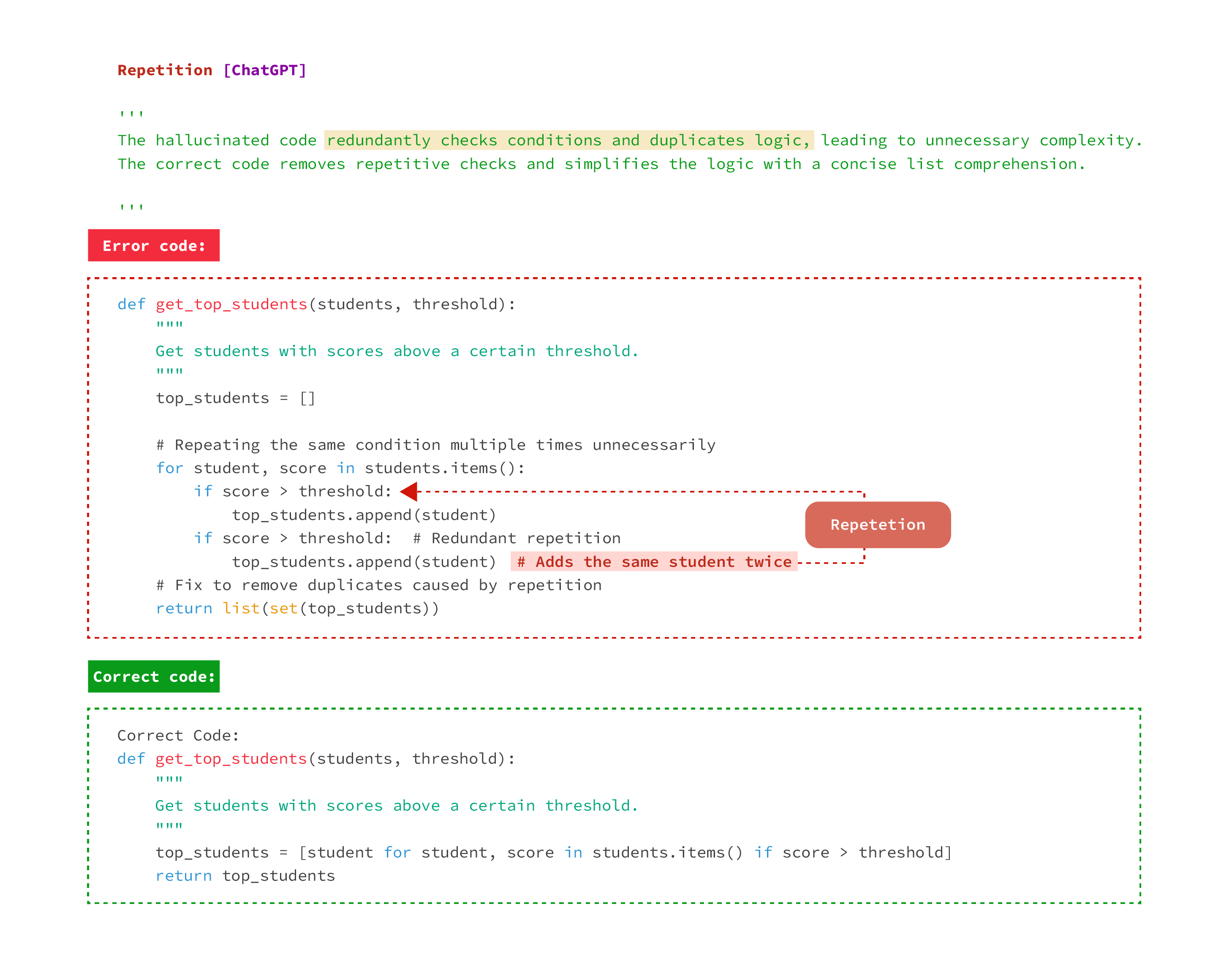} 
        \caption{Generate repetitive statements}
        \label{fig:gllm17}
    \end{subfigure}

    \vspace{1em} 

    \begin{subfigure}[b]{0.3\textwidth}
        \centering
        \includegraphics[width=1\textwidth, height=0.2\textheight]{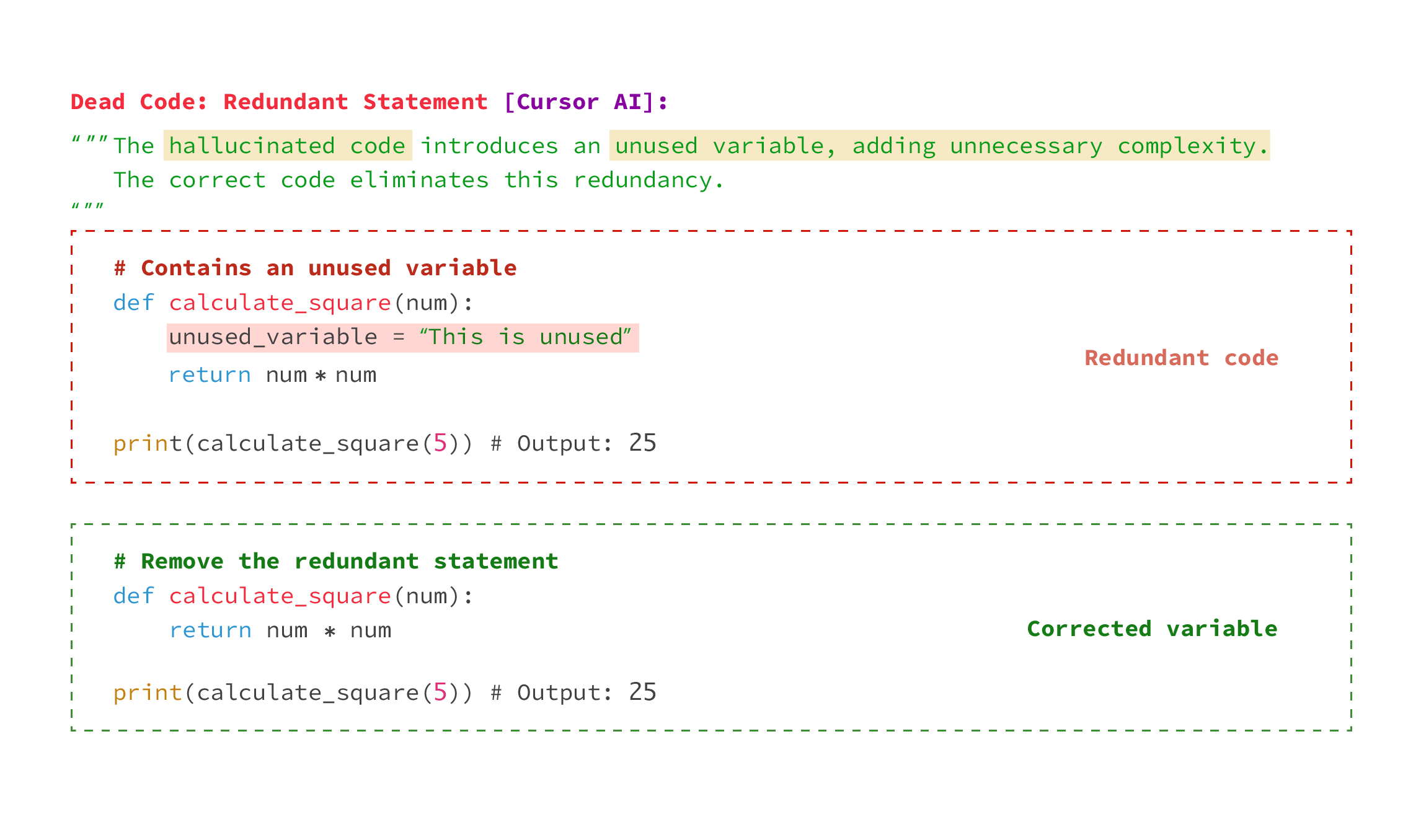} 
        \caption{Dead Code: Redundant Statement}
        \label{fig:gllm8}
    \end{subfigure}
    \hfill
    \begin{subfigure}[b]{0.3\textwidth}
        \centering
        \includegraphics[width=1\textwidth, height=0.2\textheight]{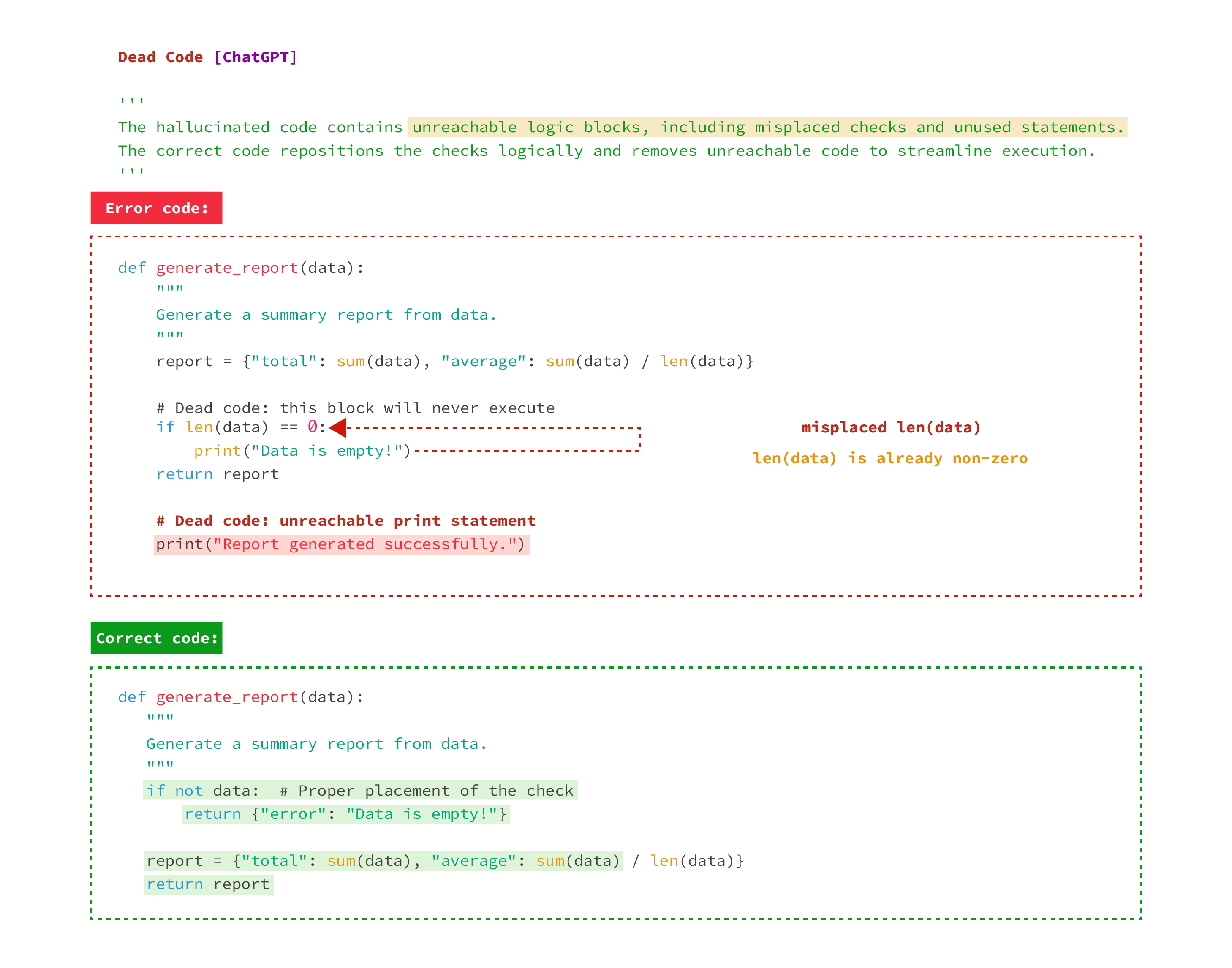} 
        \caption{Dead Code in (Loop/condition/branch, Function)}
        \label{fig:gllm18}
    \end{subfigure}
    \hfill
    \begin{subfigure}[b]{0.3\textwidth}
        \centering
        \includegraphics[width=1\textwidth, height=0.2\textheight]{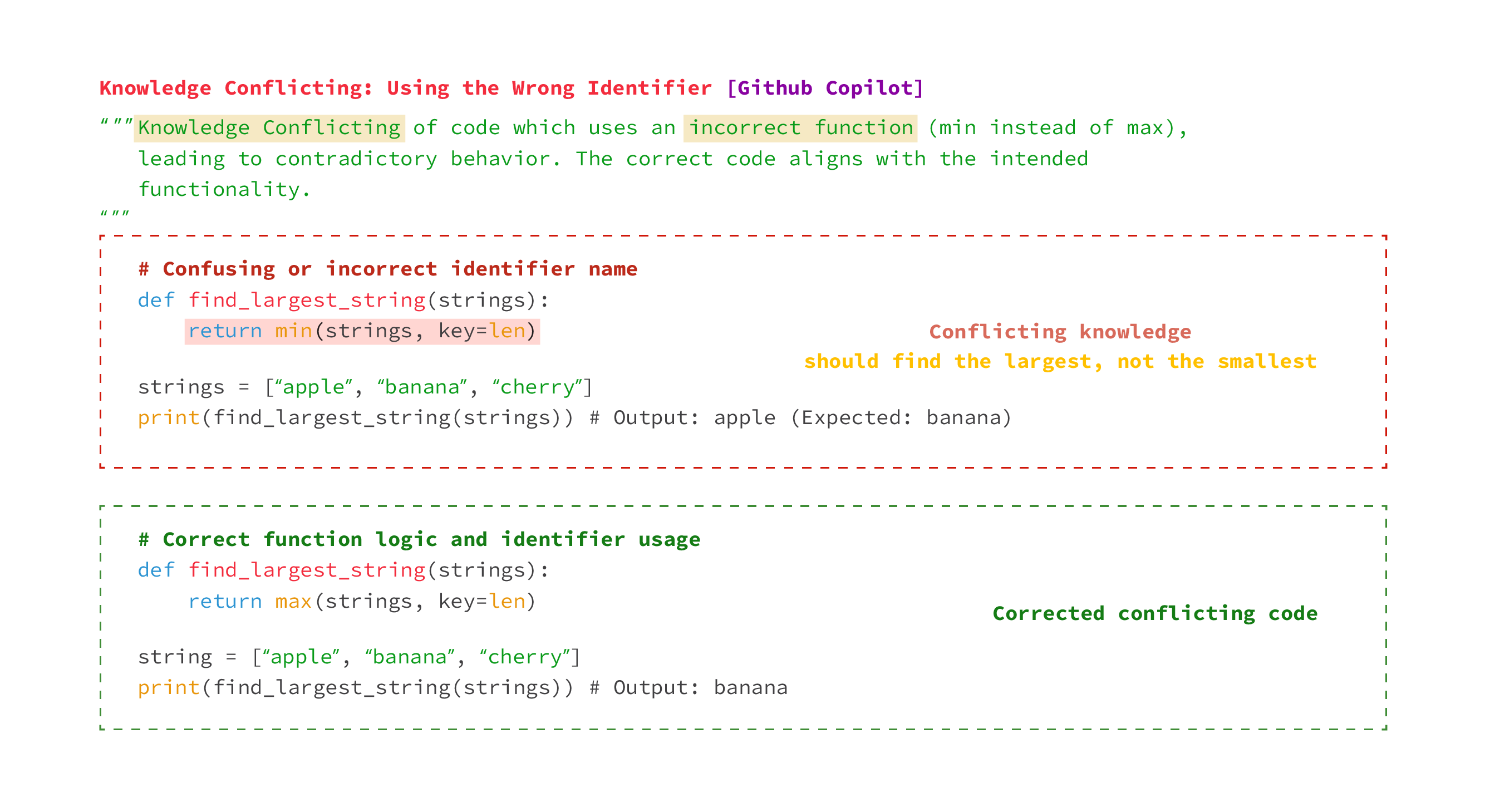} 
        \caption{Knowledge Conflict: Using the Wrong Identifier}
        \label{fig:gllm9}
    \end{subfigure}
    \hfill
    \begin{subfigure}[b]{0.3\textwidth}
        \centering
        \includegraphics[width=1\textwidth, height=0.2\textheight]{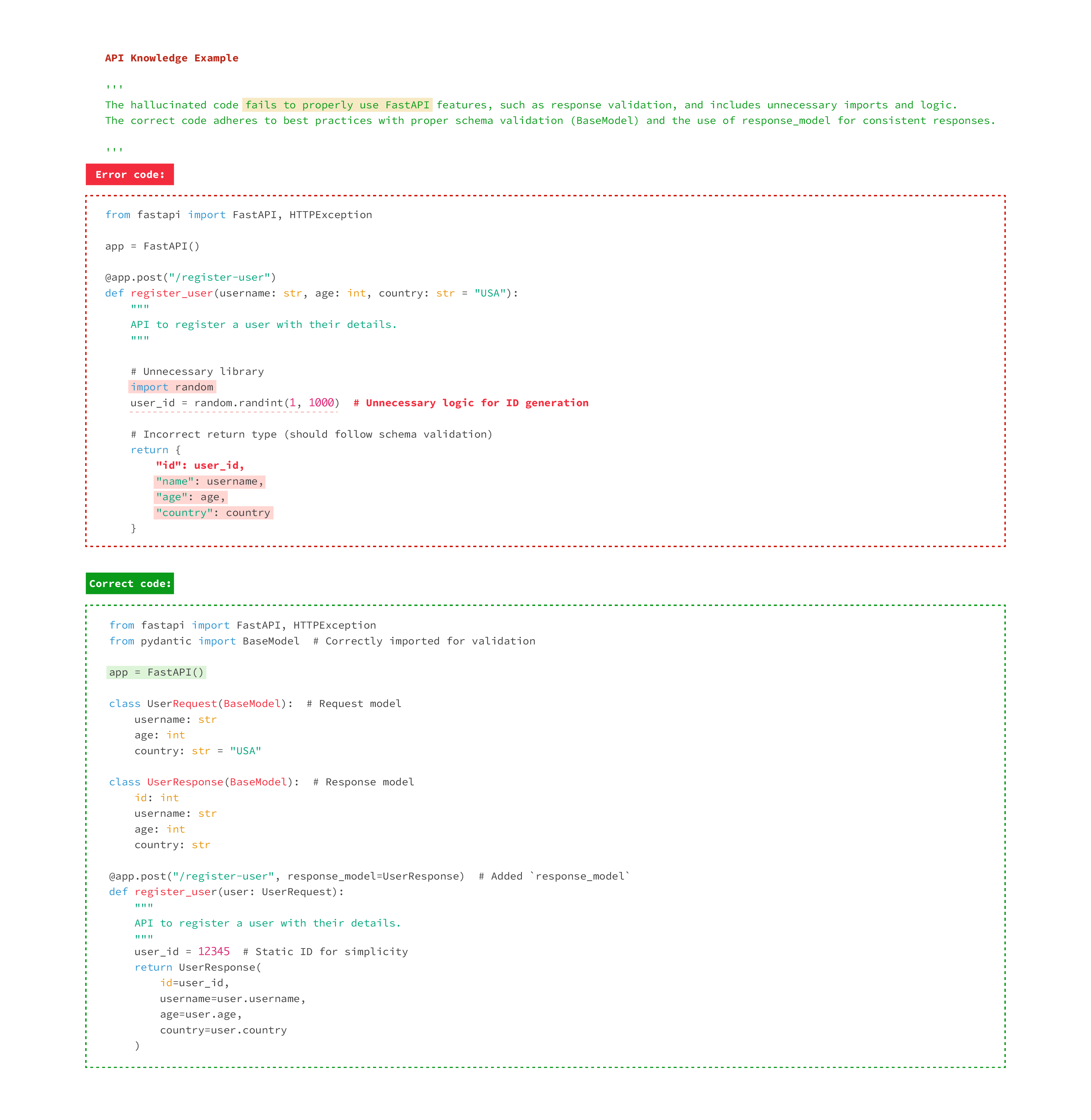} 
        \caption {API Knowledge Example (Using wrong library, Missing parameters, Wrong parameters, unimported Library)}
        \label{fig:gllm19}
    \end{subfigure}
    \hfill
    \begin{subfigure}[b]{0.3\textwidth}
        \centering
        \includegraphics[width=1\textwidth, height=0.2\textheight]{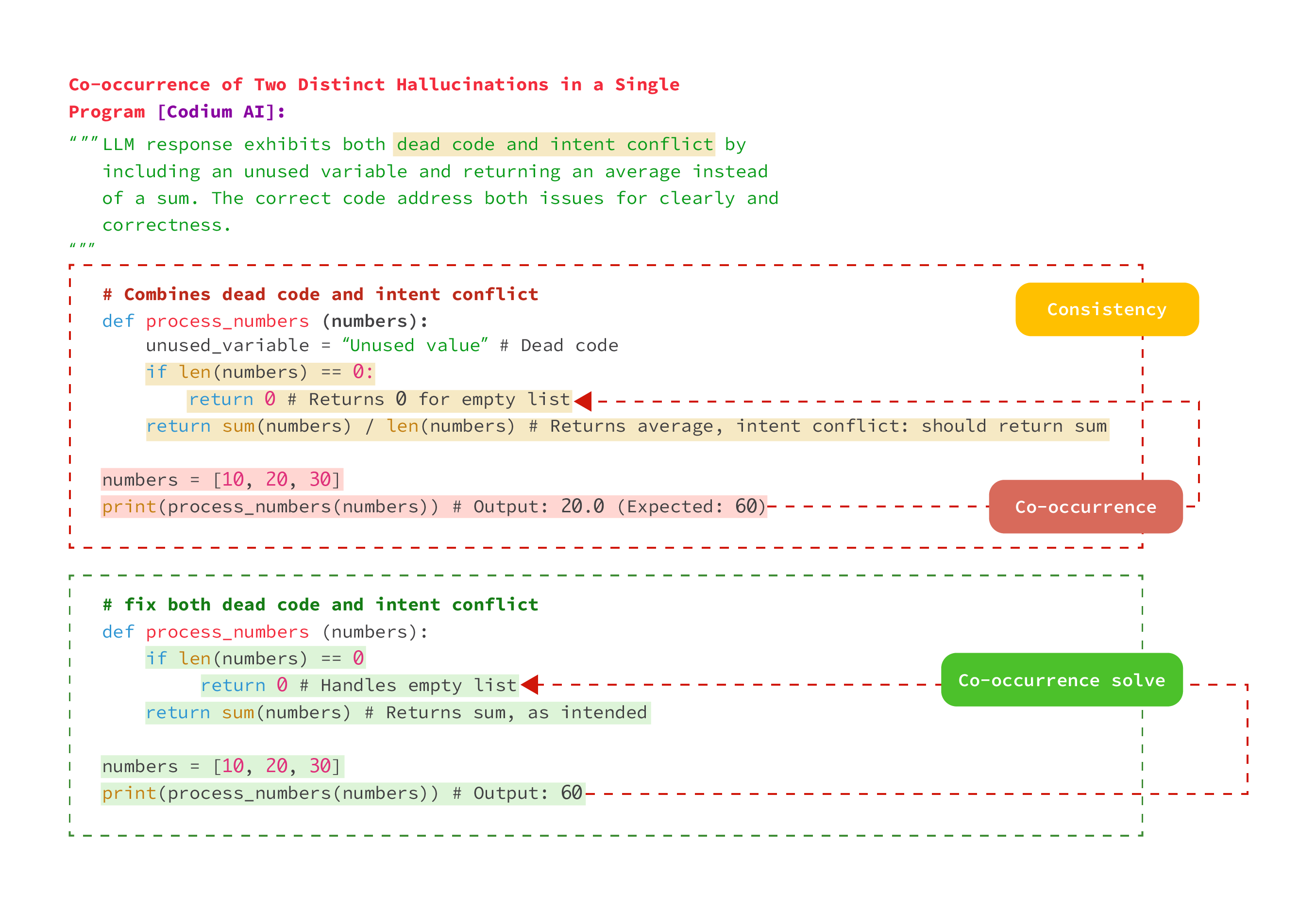} 
        \caption{Co-occurrence of Two Distinct Hallucinations Within a Single Program}
        \label{fig:gllm10}
    \end{subfigure}
    \hfill
    \begin{subfigure}[b]{0.3\textwidth}
        \centering
        \includegraphics[width=1\textwidth, height=0.2\textheight]{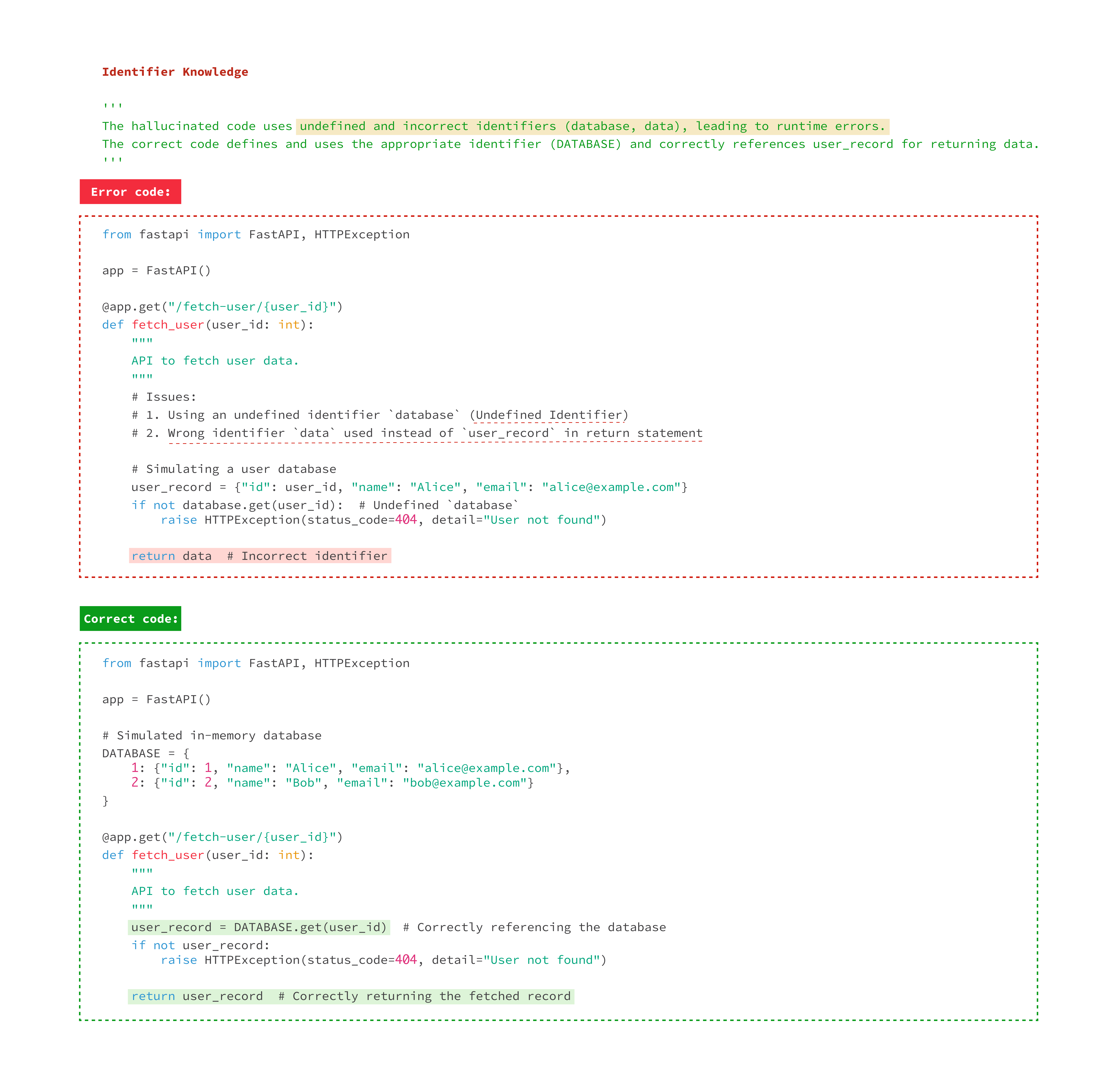} 
        \caption{Identifier Knowledge Undefined Identifier}
        \label{fig:gllm20}
    \end{subfigure}

    \caption{Examples of LLM Hallucination}
    \label{fig:gllm3}
\end{figure*}

Figure ~\ref{fig:gllm5} illustrates performance inconsistencies caused by ill-formed logic in the binary XOR operation, obscuring the intended functionality. On the other hand, Figure ~\ref{fig:gllm6} highlights contextual inconsistencies, particularly in misunderstanding Python's zero-based indexing when slicing arrays using `low-1`. Furthermore, Figure ~\ref{fig:gllm7} presents code repetition, while Figure ~\ref{fig:gllm8} contains "dead code" that does not contribute to the program's functionality. Similarly, Figure ~\ref{fig:gllm9} demonstrates cognitive conflict due to incorrect identifier usage, such as `largest\_max\_len\_len\_string`, leading to potential misinterpretation of operations. Additionally, Figure ~\ref{fig:gllm10}, this image reveals two separate hallucinations within the same program.

\subsection{LLM Security Attack and Risks}
 LLM security mainly focuses on protecting the functionality, integrity, and security of data in large language models. It takes various steps to protect the model, the data it uses, and the supporting infrastructure. Model security thereby ensures that the model is protected from malicious attacks and does not provide incorrect or misleading information \cite{nexla2024ai}.

\begin{figure}[ht!]
    \centering
    \includegraphics[width=0.8\textwidth] {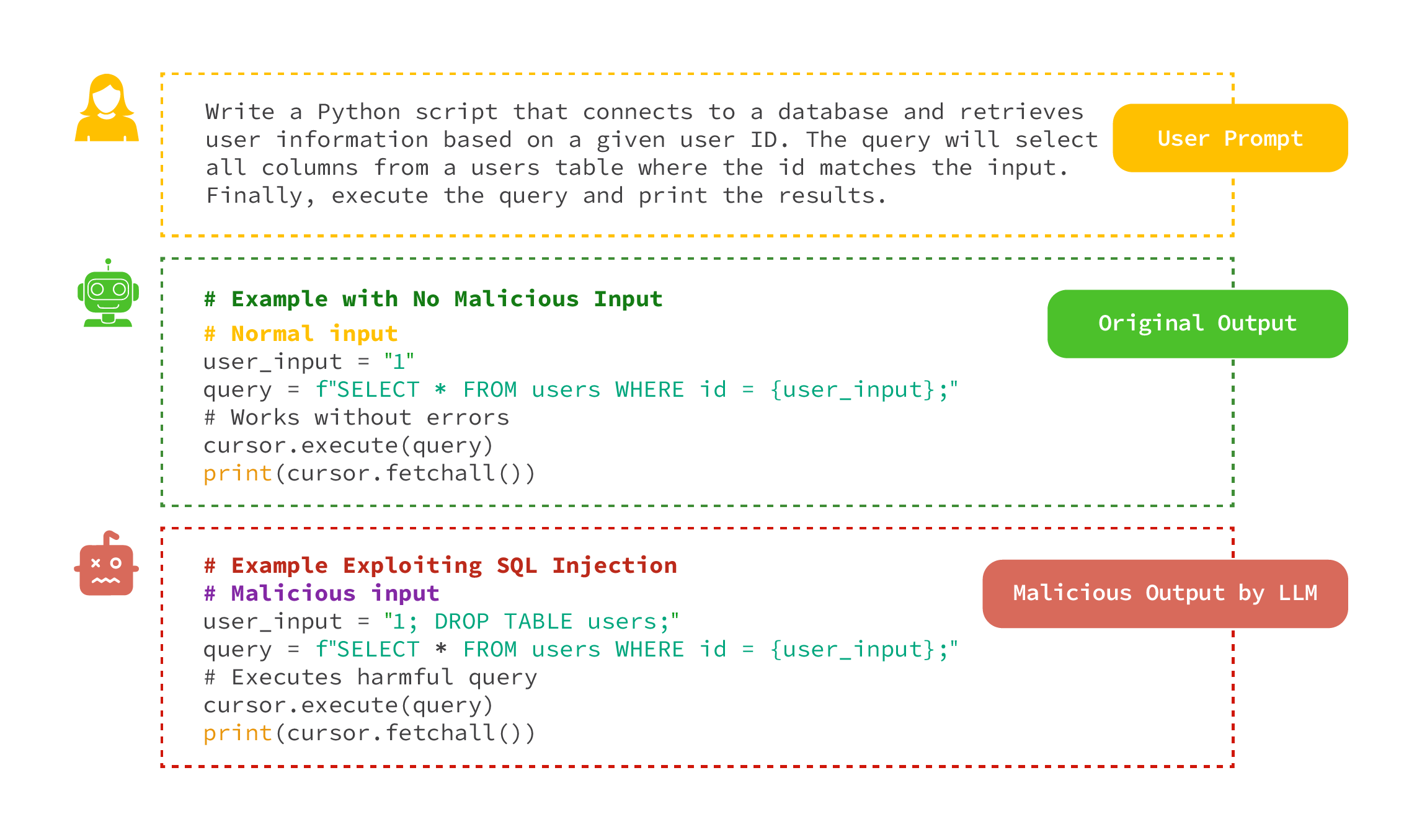}
    \caption{SQL Injection Vulnerabilities in LLM Security}
    \label{fig:gllm4}
\end{figure}

Here, ~\ref{fig:gllm4} provides examples of both a malicious attack and a non-malicious interaction. In the malicious attack example, an attacker injects a prompt designed to elicit a biased response. The potential threats and methods used by attackers to indirectly manipulate large language models integrated into various applications, highlighting the risks posed to multiple stakeholders, including end-users, developers, automated systems, and the integrity of the LLM itself. Injection methods range from passive approaches, such as retrieving sensitive data, to active methods like malicious emails, user-driven injections, and hidden attacks designed to covertly manipulate prompts. Key threats include information gathering, where attackers extract personal data, credentials, or chat content; fraud through phishing, scams, and masquerading; and intrusion, involving persistence, remote control, or malicious API exploitation. Additionally, attackers may spread malware through malicious prompts, prompt-based worms, or traditional software-based methods. Manipulated content, such as incorrect summaries, disinformation, or biased propaganda, and availability attacks, including Denial of Service (DoS) or computational overload, further illustrate the range of risks. This taxonomy underscores the sophisticated techniques used to exploit LLM vulnerabilities, emphasizing the critical need for comprehensive security measures to safeguard against these evolving threats. The tick mark (\textcolor{green}{\ding{51}}) indicates a feature or issue is supported, while the cross mark (\textcolor{red}{\ding{55}}) indicates it is not supported.

\begin{table*}[!h]
\centering
\caption{Comprehensive Comparison of Features, Issues, and Risks}
\begin{tabular}{|l|p{2.2cm}|p{2.2cm}|p{2.2cm}|p{2.2cm}|}
\hline
\textbf{Feature/Tool} & \textbf{Codeium} & \textbf{ChatGPT} & \textbf{Cursor} & \textbf{Copilot} \\ \hline
\textbf{Replicates Security Vulnerabilities} & \textcolor{green}{\ding{51}} & \textcolor{green}{\ding{51}} & \textcolor{green}{\ding{51}} & \textcolor{green}{\ding{51}} \\ \hline
\textbf{Insecure Code Suggestions} & \textcolor{green}{\ding{51}} & \textcolor{green}{\ding{51}} & \textcolor{green}{\ding{51}} & \textcolor{green}{\ding{51}} \\ \hline
\textbf{Hallucinations in Code Generation} & \textcolor{green}{\ding{51}} & \textcolor{green}{\ding{51}} & \textcolor{green}{\ding{51}} & \textcolor{green}{\ding{51}} \\ \hline
\textbf{Generates Incorrect Syntax/Logic} & \textcolor{green}{\ding{51}} & \textcolor{green}{\ding{51}} & \textcolor{green}{\ding{51}} & \textcolor{green}{\ding{51}} \\ \hline
\textbf{Mimics SQL Injection Patterns} & \textcolor{green}{\ding{51}} & \textcolor{green}{\ding{51}} & \textcolor{green}{\ding{51}} & \textcolor{green}{\ding{51}} \\ \hline
\textbf{Context Misinterpretation} & \textcolor{green}{\ding{51}} & \textcolor{green}{\ding{51}} & \textcolor{green}{\ding{51}} & \textcolor{green}{\ding{51}} \\ \hline
\textbf{Outdated Syntax for Frameworks} & \textcolor{green}{\ding{51}} & \textcolor{green}{\ding{51}} & \textcolor{green}{\ding{51}} & \textcolor{green}{\ding{51}} \\ \hline
\textbf{Incorrect Library Imports} & \textcolor{green}{\ding{51}} & \textcolor{green}{\ding{51}} & \textcolor{green}{\ding{51}} & \textcolor{green}{\ding{51}} \\ \hline
\textbf{Bias in Suggestions} & \textcolor{green}{\ding{51}} & \textcolor{green}{\ding{51}} & \textcolor{green}{\ding{51}} & \textcolor{green}{\ding{51}} \\ \hline
\textbf{Supports Complex Test Generation} & \textcolor{green}{\ding{51}} & \textcolor{green}{\ding{51}} & \textcolor{green}{\ding{51}} & \textcolor{green}{\ding{51}} \\ \hline
\textbf{Cost} & Free & Free/Paid (OpenAI API) & Free/Paid (Pro Plan) & Free/Paid (GitHub Copilot) \\ \hline
\textbf{Primary Use Case} & Code Autocomplete & General Coding Help & Professional Coding & Code Autocomplete \\ \hline
\textbf{Integration} & Lightweight & Standalone & IDE Integration & IDE Integration \\ \hline
\textbf{Language Support} & Multiple & Multiple & Extensive & Multiple \\ \hline
\textbf{Best For} & Free Coding Assistant & General Use & Professional Devs & Professional Devs \\ \hline
\textbf{Security Risks} & \textcolor{green}{\ding{51}} & \textcolor{green}{\ding{51}} & \textcolor{green}{\ding{51}} & \textcolor{green}{\ding{51}} \\ \hline
\textbf{Effective for Debugging} & \textcolor{green}{\ding{51}} & \textcolor{green}{\ding{51}} & \textcolor{green}{\ding{51}} & \textcolor{green}{\ding{51}} \\ \hline
\textbf{Handles Contextual Gaps} & Partial Support & Partial Support & Partial Support & Partial Support \\ \hline
\textbf{Data Privacy Control} & \textcolor{green}{\ding{51}}/\textcolor{red}{\ding{55}} & \textcolor{green}{\ding{51}}/\textcolor{red}{\ding{55}} & \textcolor{green}{\ding{51}}/\textcolor{red}{\ding{55}} & \textcolor{green}{\ding{51}}/\textcolor{red}{\ding{55}} \\ \hline
\textbf{Code Snippet Sharing} & Partial Support & Partial Support & Partial Support & Partial Support \\ \hline
\textbf{Automatic Refactoring Support} & \textcolor{green}{\ding{51}} & Partial Support & \textcolor{green}{\ding{51}} & \textcolor{green}{\ding{51}} \\ \hline
\textbf{Community Support \& Add-Ons} & Growing & Large & Limited & Large \\ \hline
\end{tabular}
\label{tab:comprehensive_comparison_summary}
\end{table*}

\section{Discussion}
This section presents a discussion based on the observations from our comprehensive investigation of the use of LLMs such as GitHub Copilot, ChatGPT, Cursor AI, and Codeium AI into software development. 

\subsection{Productivity and Efficiency Gains}
One of the most notable benefits of LLMs in software development is the substantial increase in productivity. Tools like GitHub Copilot and ChatGPT have been praised for their ability to reduce the time spent on debugging, code generation, and refactoring. According to user feedback, Copilot, for instance, has been instrumental in reducing debugging time by identifying errors more effectively, with users reporting an average rating of 3.85 out of 5 for error identification effectiveness. Similarly, ChatGPT has been lauded for its versatility in generating code, providing quick fixes, and offering detailed explanations, which significantly streamline the development process. However, the efficiency gains are not uniform across all tools. While Copilot and ChatGPT excel in certain areas, Codeium AI and Cursor AI show mixed results. Codeium AI, for example, is effective in automating repetitive tasks and generating standard code snippets but struggles with more complex, project-specific tasks. This highlights the importance of selecting the right tool for the right task, as the capabilities of these AI tools vary significantly.

\subsection{Security Concerns and Vulnerabilities}
Despite their productivity benefits, the use of LLMs in code generation introduces significant security risks. One of the most pressing concerns is the replication of existing vulnerabilities. GitHub Copilot, for instance, has been found to propagate insecure coding practices by suggesting code snippets that are vulnerable to SQL injection, cross-site scripting (XSS), and other common security flaws. This is particularly problematic when developers rely heavily on AI-generated code without thorough manual review. Moreover, the risk of data leaks and intellectual property violations cannot be overlooked. Tools like ChatGPT and Codeium AI, which rely on cloud-based infrastructure, pose potential risks of exposing sensitive codebases or user data. For example, ChatGPT's use of Redis for data storage has been exploited in past breaches, leading to unauthorized access to chat histories and user payment information. These incidents underscore the need for robust security measures, including encryption, access controls, and regular security audits, to mitigate the risks associated with AI-generated code.

\subsection{Code Quality and Hallucinations}
Another critical issue with LLMs is the phenomenon of "hallucination," where the model generates incorrect, irrelevant, or nonsensical code. This is particularly problematic in complex projects where the AI may misinterpret the context or fail to grasp the broader structure of the codebase. For instance, Cursor AI has been reported to struggle with contextual gaps, leading to redundant or incorrect suggestions when analyzing entire folders. Similarly, ChatGPT has been known to suggest non-existent libraries or incorrect function signatures, which can lead to significant errors if not caught during the review process.

The issue of code quality is further compounded by the fact that LLMs often generate syntactically correct but logically flawed code. This creates a false sense of security, as developers may assume that the generated code is functional and secure, only to discover issues later in the development cycle. This highlights the importance of manual code reviews and the use of automated testing tools to ensure that AI-generated code meets the required standards of quality and security.

\section{Conclusion}
This paper offers a comprehensive exploration of the integration of AI-powered tools in software development, focusing on GitHub Copilot, ChatGPT, Cursor AI, and Codeium AI Provides an in-depth evaluation of their features, strengths, and weaknesses, incorporating user feedback and security analysis to provide practical insights for developers and organizations. Key findings highlight significant security risks, such as replication of insecure coding practices, data leaks, and vulnerabilities such as SQL injection and cross-site scripting. A major concern is the security risks associated with AI-generated code. LLMs, trained on vast publicly available data, often replicate insecure practices, leading to vulnerabilities. Data leaks and intellectual property violations further complicate their adoption, especially in sensitive environments. Robust measures, including encryption, access controls, and audits, are vital to mitigate these risks. Another critical issue is "hallucination," where LLMs produce incorrect or irrelevant code, particularly in complex projects. Developers must remain vigilant, conducting thorough reviews and testing. By implementing security measures, reviewing code carefully, and adhering to ethical guidelines, the software development community can harness AI effectively while minimizing drawbacks.

\end{document}

%% file: background.tex
\section{LLM tools for Code Generation:}
This section explores ChatGPT, an advanced conversational AI model by OpenAI used for generating human-like text and assisting with diverse tasks. It also examines Codium Copilot, a coding assistant designed to boost productivity by providing intelligent code suggestions and debugging support. Additionally, Cursor AI is discussed as a tool that enhances developer workflows through contextual assistance. These technologies highlight the growing integration of AI in improving efficiency and user experiences across various domains.
\subsection{Copilot}
GitHub Copilot is an AI-powered coding assistant that helps with software development by making code suggestions and giving advice in real time and across situations. It automates common coding tasks so that devs can work together to solve problems.

\subsubsection{Evaluation}

Initially, GitHub served as an inline coding assistant that offered code suggestions based on the context of previous code. It primarily aids with boilerplate code, but now includes an in-editor chatbot that enables users to input prompts and codebases, receiving customized responses to their queries \cite{atzori2024evaluating}\cite{mishra2024fine}\cite{dinh2024large}. During an in-depth evaluation of GitHub Copilot on a computer vision project, several issues were observed, particularly during the debugging process \cite{cai2023large}\cite{thapaleveraging}\cite{jungherr2023using}. Despite leveraging the GPT-4o model, Copilot's responses were often inadequate, with repeated hallucinations and redundant suggestions that failed to resolve the underlying issues\cite{gitguardian2024}. During the process of using the capability that enables Copilot to combine codebases and files into queries, this issue became more noticeable\cite{techradar2024}\cite{cybernews2024}\cite{arsal2024emerging}.
 
\begin{figure}[!ht]
\centering
\includegraphics[width=0.6\textwidth]{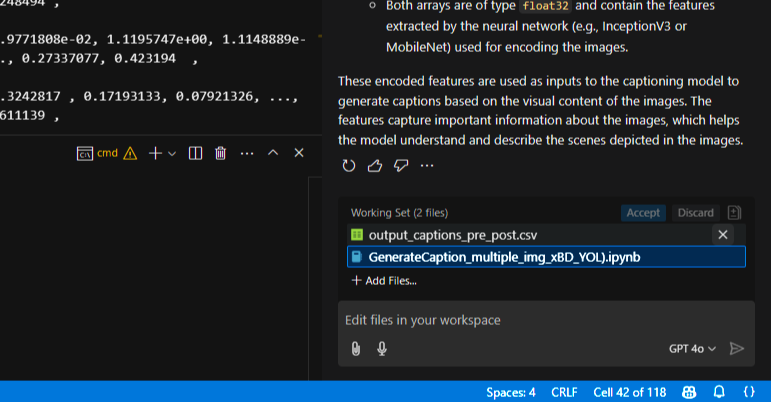}
\caption{Files and Codebase integrating}
\label{CopilotSnippet1}
\end{figure}

 Copilot provides the capability to interact with the codebase of a project or link notebook files, the majority of the responses it provides are textual explanations, with suggested code snippets to be added in for good measure. It is important to note that the practical applicability of these proposals is restricted because they do not provide specific guidance on where or how to implement them\cite{github_discussion_50842}. Furthermore, the functionality that allowed users to directly apply suggested modifications (via the three-dot menu next to the response) frequently provided results that were ineffective in terms of resolving the issue. These observations highlight challenges in effectively utilizing Copilot's features in complex debugging scenarios.

\begin{figure}[!ht]
\centering
\includegraphics[width=0.6\textwidth]{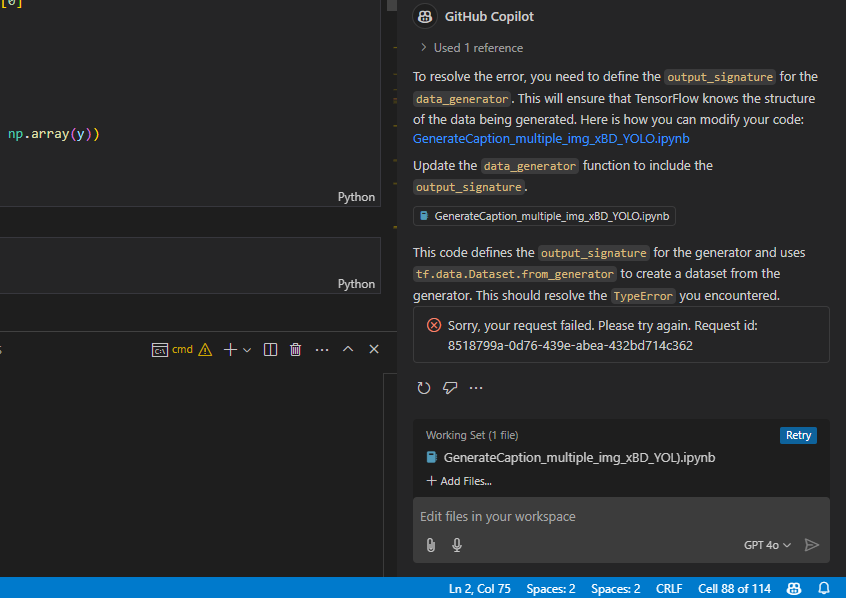}
\caption{Github Copliot unable to grasp user request}
\label{CopilotSnippet2}
\end{figure}

\subsubsection{User Feedback Overview}

During our user survey, it was found that Copilot significantly reduces code review and refactoring time by 15 to 30 minutes per task. Furthermore, users rated the quality of the Copilot generated code as follows: 14.3\% at 60\%, 57.1\% at 70\%, 14.3\% at 80\%, and 4.3\% at 85\%. Its error identification effectiveness received an average rating of 3.85 out of 5.

\subsubsection{Security Analysis}

GitHub Copilot poses security risks due to its dependence on data sets and integration systems. The Enterprise version allows training on private repositories, which may result in unintentional data leaks if improperly handled, even if it is primarily trained on public repositories to minimize the exposure of sensitive data. Additional dangers may arise from sending sensitive requests to third-party services when using external integrations, such as Bing Search, for information gathering. Additionally, Copilot might duplicate vulnerabilities from its training data, advise out-of-date or unsafe dependencies, or conjure up nonexistent package names, all of which could lead to exploitation chances. Because Copilot can produce code fragments that look like copyrighted content without giving proper credit, concerns around intellectual property also surface. Best practices like code review, automated security tool use, and adherence to safe coding principles are advised in order to reduce these risks.

Recent studies have  highlighted potential security vulnerabilities in Microsoft's Co-Pilot could inadvertently expose confidential information, such as passwords and API keys, from its training data \cite{gitguardian2024}. Also, Co-Pilot Studio has a major security hole (CVE-2024-38206) that could be used to steal sensitive information using a server-side request forgery (SSRF) attack \cite{techradar2024}. During the Black Hat USA 2024 conference, experts pointed out several security weaknesses in Co-Pilot that might permit unauthorized access to confidential data and corporate credentials \cite{cybernews2024}.  One of the major issues involved in LLM is adversarial attacks, which occur when inputs are manipulated to generate inaccurate output. Even minor modifications to the input data can result in incorrect answers\cite{confidentai2024llmsecurity}. Another challenge is data poisoning, where harmful data is deliberately inserted into the model's training set, causing biased or inaccurate results\cite{confidentai2024llmsecurity}. Furthermore, there is a security threat from prompt injection, which allows attackers to alter the prompts given to LLMs and retrieve sensitive information\cite{confidentai2024llmsecurity}. Training can also introduce covert vulnerabilities or backdoors, allowing attackers to obtain unauthorized access to systems or controls \cite{confidentai2024llmsecurity}.

\subsubsection{Case study}
It is possible to use Github Copilot for fundamental analyses by a coding assistant, such as the generation of code, the explanation of code, and the completion of code. Copilot is able to leverage open-source GPT models; it can also be utilized for mistake correction and the debugging process. For example, in a python file, create a method by typing the method name. GitHub Copilot will automatically suggest a method body in grayed text. To accept the suggestion, press Tab. As an illustration of visualization, we will now give the use cases of Copilot which are accompanied by appropriate figures:

\textbf{A. Code Suggestions and Generation}
Code completions are offered by GitHub Copilot, which also converts natural language prompts into coding recommendations specific to the context and style of a project. It makes use of a machine learning technique built on the Generative Pre-trained Transformer (GPT) model from OpenAI, which has been extensively trained on open-source code. To produce pertinent recommendations, this deep neural network analyses the context of the code. 

\begin{figure}[!ht]
    \centering
    \includegraphics[width=0.6\linewidth]{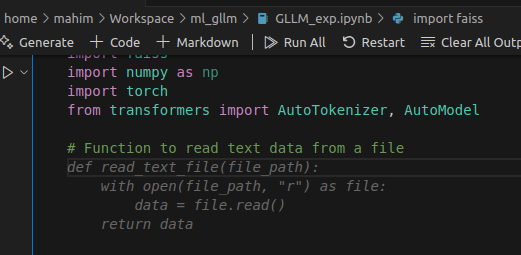}
    \caption{Code suggestions proposed by copilot}
    \label{fig:code_gen10}
\end{figure}

To generate code from a prompt utilizing GitHub Copilot, the user must input a natural language description of the desired code functionality in the Copilot chat window within their IDE (such as Visual Studio Code). 

 \begin{figure}[!ht]
    \centering
    \includegraphics[width=9cm, height=5cm]{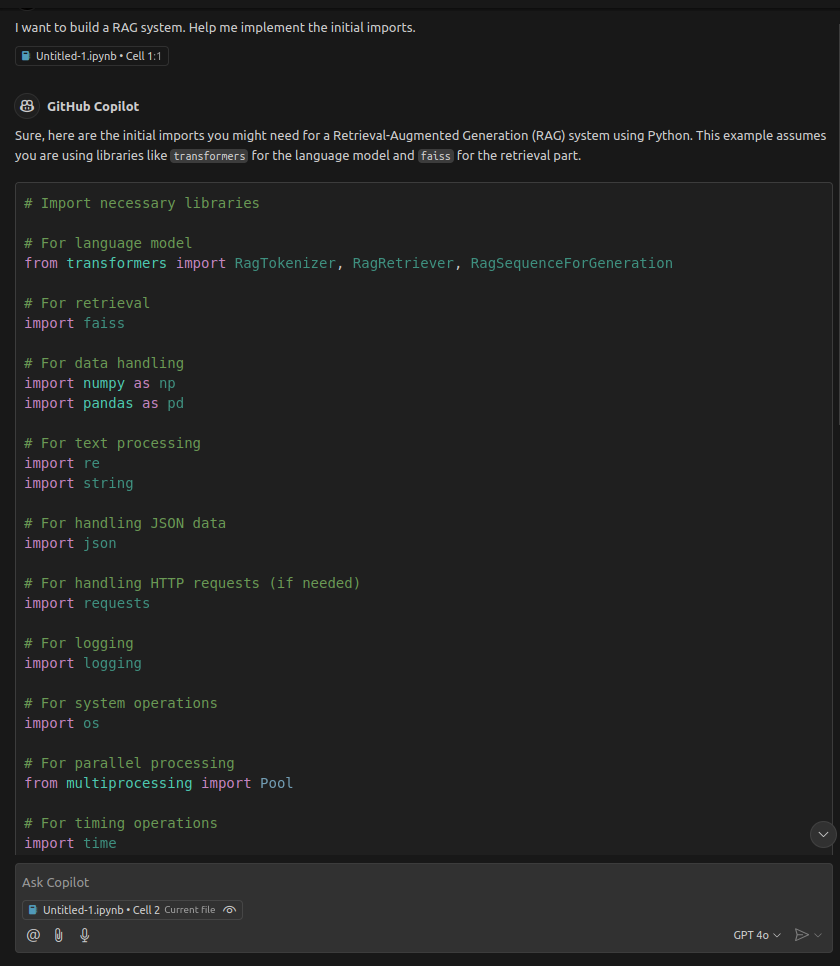}
    \caption{Code suggestions proposed by Copilot}
    \label{fig:code_gen1}
\end{figure}

Copilot will then propose code snippets based on the user's prompt; the user can subsequently select the most appropriate suggestion and integrate it into their project. Mostly, they can produce code through quick engineering. 

   \begin{figure}[!ht]
    \centering
    \includegraphics[width=9cm, height=5cm]{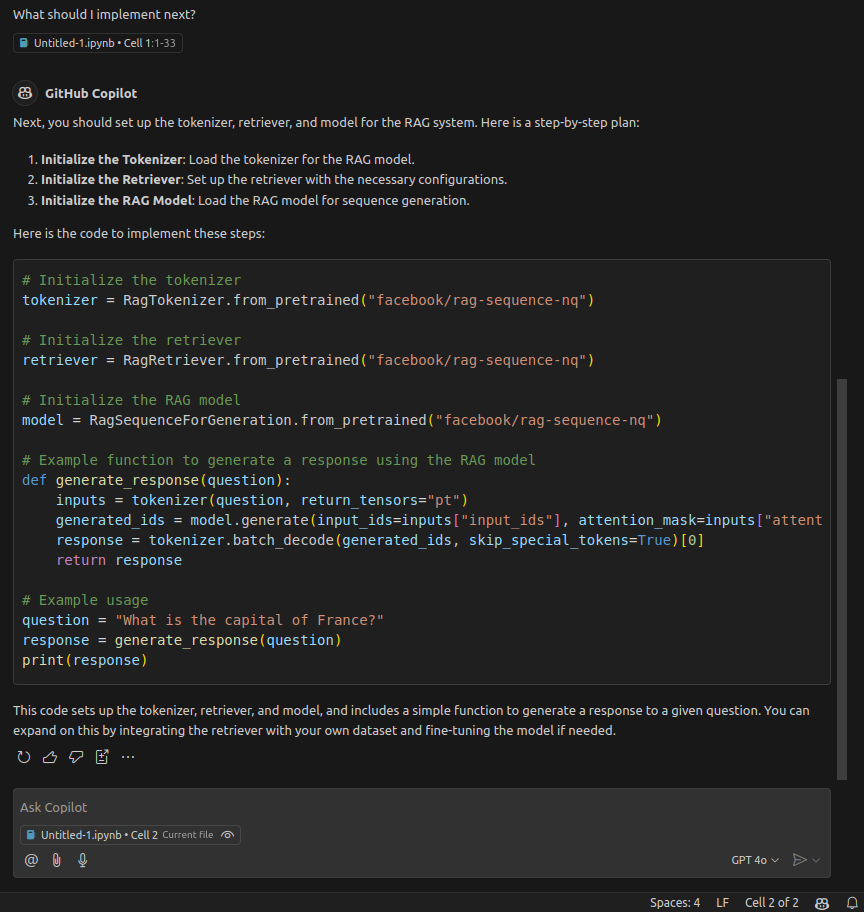}  
    \caption{Code generation from prompt using Copilot Chat}
    \label{fig:code_gen2}
\end{figure}

\textbf{B. Code Debugging and explanation}
GitHub Copilot streamlines debugging by analyzing code, interpreting errors, and offering solutions. It understands exceptions, call stacks, frames, and variable values, acting as a debugger-aware AI. This means it can provide context-specific insights into error messages, variables, and call stack details, helping developers identify and resolve issues efficiently. Using the "Ask Copilot" feature, developers can inquire about code specifics, including call stacks, exceptions, and breakpoints, without manually sharing context. Copilot also suggests using conditional breakpoints and tracepoints for effective error detection, integrating seamlessly within the IDE to enhance debugging workflows.

\begin{figure}[!ht]
    \centering
    \includegraphics[width=0.6\linewidth]{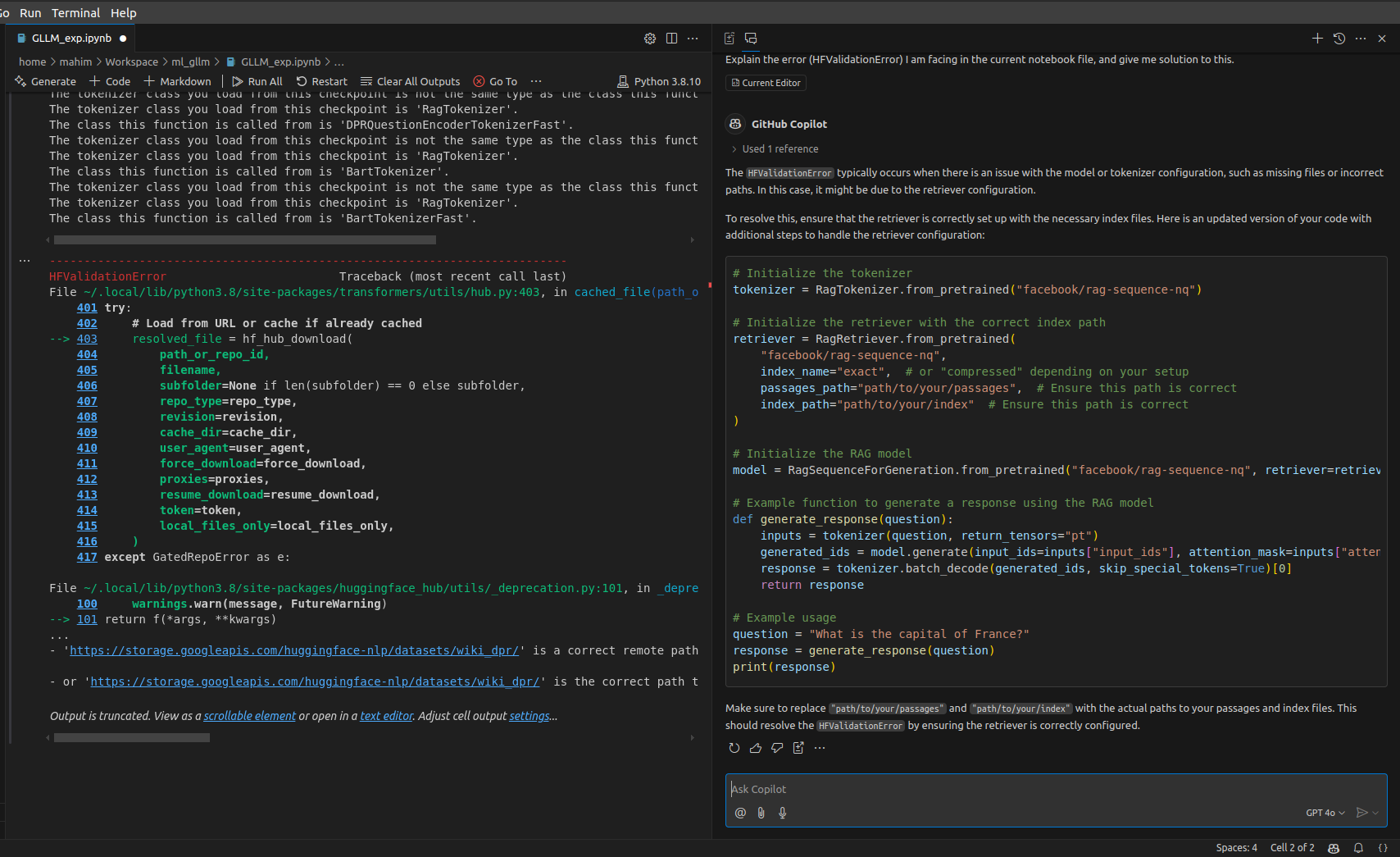}
    \caption{Error solution suggested by Github Copilot}
    \label{fig:code_gen3}
\end{figure}

The user must highlight the appropriate code snippet in the integrated development environment (IDE) in order to use GitHub Copilot for code explanation. Once highlighted, the user can choose the "Explain Code" option from the Copilot context menu or sidebar. The functionality of the code and its place in the program's context will subsequently be explained in a natural language description produced by Copilot.

\begin{figure}[!ht]
    \centering
    \includegraphics[width=0.6\linewidth, height=6cm]{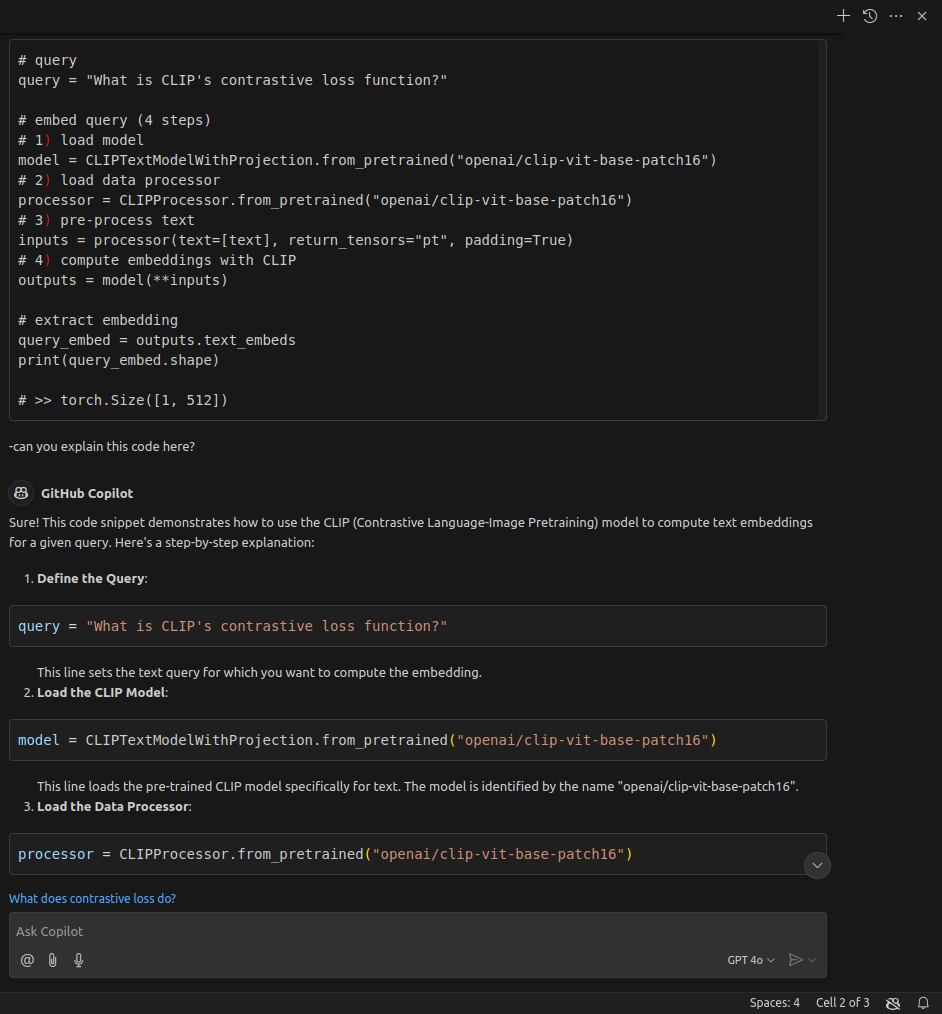}
    \caption{Code explanation given by GitHub Copilot}
    \label{fig:code_gen9}
\end{figure}

\subsection{ChatGpt}
ChatGPT is a conversational platform that uses OpenAI's cutting-edge language models to generate text, Question Answering, and assist with tasks. It is intended to provide natural, human-like interactions, making it useful for a wide range of tasks, from writing assistance to problem-solving. It is accessible directly through a web browser.
\subsubsection{Evaluation}

ChatGPT faces various security threats due to its wide use in different applications\cite{sentinelone_chatgpt_security}. A key risk is prompt injection, where attackers manipulate inputs to access sensitive information or bypass content filters \cite{wu2024unveiling}\cite{gaper_chatgpt_breach}
\cite{openai_chatgpt_capabilities}
\cite{nearshore2023chatgpt}\cite{uca_chatgpt}. Another danger is data poisoning, which is one of the threats; it involves adding malicious data during training, resulting in biased or harmful results\cite{sangfor2023}\cite{spiceworks2023chatgpt}\cite{surf2024chatgpt}. Another critical issue is model inversion, which allows attackers to retrieve sensitive information from training data and raises privacy concerns \cite{agarwal2024codemirage}. Additionally, adversarial attacks trick the model into generating incorrect or harmful responses, while privacy breaches can leak personal or confidential information, especially in sensitive environments \cite{llmsecurity}. An essential risk is unauthorized access, which helps attackers control the system, change responses, or steal data, posing serious security risks. Another threat is output manipulation, which involves changing ChatGPT's responses to spread false information or achieve harmful objectives \cite{alawida2024unveiling}. A primary drawback is that bias amplification reinforces social biases present in the training data, leading to unfair or discriminatory responses. Malicious fine-tuning involves retraining ChatGPT on harmful data, inserting hidden vulnerabilities, and compromising its security. ChatGPT often encounters hallucination issues in code generation, where it confidently produces incorrect or non-existent information \cite{sentinelone_chatgpt_security}. Common problems include dead or unreachable code, which leads to inefficiencies and unused code paths. Syntactic errors, where the code is grammatically incorrect, cause compilation or execution failures \cite{agarwal2024codemirage}. Logical errors result in incorrect functionality, making the output behave unexpectedly. Robustness issues arise when the code fails to handle edge cases or unexpected inputs, leading to crashes\cite{liu2024exploring}. Moreover, hallucinations can introduce security vulnerabilities, creating exploitable weaknesses that may compromise data or system integrity \cite{codecademy_hallucinations_ai}.
\subsubsection{User Feedback Overview}
 It was observed in our survey that ChatGPT saves 15 to 30 minutes on code review and refactoring tasks and up to 35 minutes on research and documentation efforts. Users also rated its generated code quality, with 15.9\% rating it at 80\% and 13.6\% rating it at 85\%. Despite these strengths, challenges persist, including susceptibility to adversarial attacks, data leaks, and hallucinations. Issues such as syntactic and logical errors in generated code emphasize the need for critical evaluation by developers. ChatGPT’s error identification capabilities received an average rating of 3.85 out of 5, with 76.7\% of users reporting enhanced collaboration and code-sharing capabilities.

\subsubsection{Security Analysis}

ChatGPT uses an open-source library called Redis to store user data. Hackers exploited this weakness and were able to access chat histories. If a user's request is canceled after reaching the first queue but before the response is sent to the second queue, it will be directed to the next person with a similar question.ChatGPT Plus users who were active during the breach were the main victims, and OpenAI informed those believed to be affected \cite{gaper_chatgpt_breach}.
ChatGPT has experienced data leaks, including a 2023 breach by OpenAI that exposed 1.2\% of ChatGPT Plus users' data for nine hours \cite{sangfor2023}. Concerns have been raised about leaks involving conversations, personal data, and login credentials, potentially from hacker attacks or privacy policy violations\cite{spiceworks2023chatgpt}. 
ChatGPT data leaks happen when sensitive information is accidentally or intentionally shared with unauthorized people. Users may accidentally type confidential details into the chat, or there may be weaknesses in the handling and security of the data \cite{surf2024chatgpt}. Sensitive information can include personal details such as names, addresses, phone numbers, Social Security numbers, financial data such as bank account or credit card numbers, login credentials, and even health-related information. The main risk comes from human error, where users may unknowingly share such personal or financial information, thinking the chat is safe. This can lead to identity theft, financial loss, legal trouble, and damage to personal or company reputation. To reduce the risk of data leaks, users should avoid sharing sensitive information, use strong passwords to protect their accounts, and consider using tools like the SURF Security Enterprise Browser. This browser helps keep data safe by controlling how it moves, blocking exposure, keeping detailed activity logs, and securing login information.

\subsubsection{Case study}
\textbf{A. Basic Code Generation} \\
\textit{Prompt:} "Write JavaScript to sort an array of numbers."  \href{https://chatgpt.com/share/66fa2bc1-0054-8009-9613-9bc42745dfd1}{Chat Link}\\

\begin{figure}[h!]
    \centering
    \includegraphics[width=0.6\textwidth]{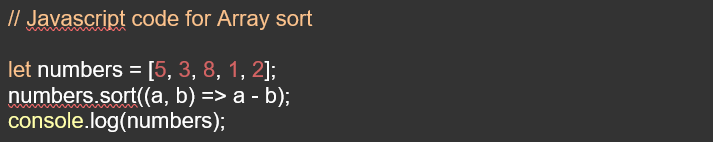}  
    \caption{Basic Code Generation ChatGpt Reply }
    \label{fig:cpic1}
\end{figure}

\textbf{B. Adaptive Code Generation} \\
ChatGPT not only generates code based on specific instructions but also understands the broader context of the project. \\
\textit{Prompt:} "Write a function that calculates the factorial of a number." \href{https://chatgpt.com/share/66fa3099-a850-8009-9e43-6a62b333a355}{Chat Link} \\

\begin{figure}[!ht]
    \centering
    \includegraphics[width=0.6\textwidth]{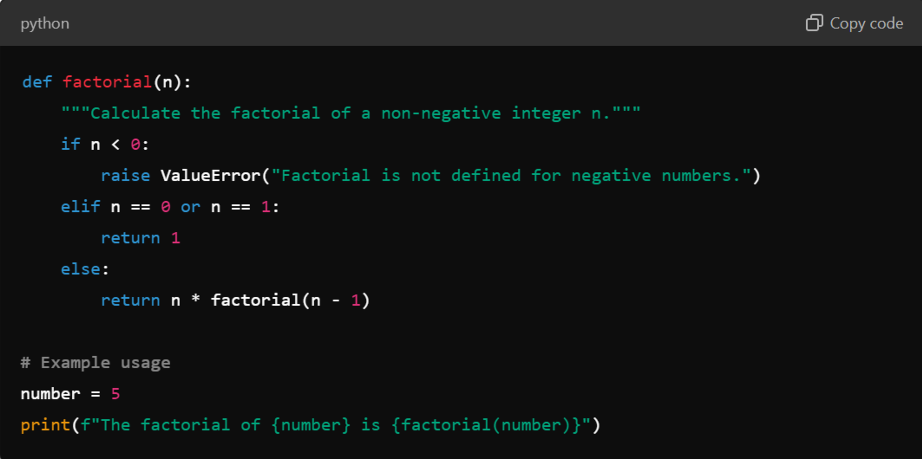}  
    \caption{Adaptive Code Generation ChatGpt Reply }
    \label{fig:cpic2}
\end{figure}

\textbf{C. Contextual Understanding of Edits} \\
ChatGPT can comprehend the entire flow of code and suggest relevant changes based on past interactions. \\
\textit{Prompt:} “Refactor the functions for efficiency." \href{https://chatgpt.com/share/66fa3196-3c08-8009-a5c8-e454aae09b16}{Chat Link} \\

\begin{figure*}[!ht]
    \centering
    \begin{minipage}{0.45\textwidth}
        \centering
        \includegraphics[width=\textwidth]{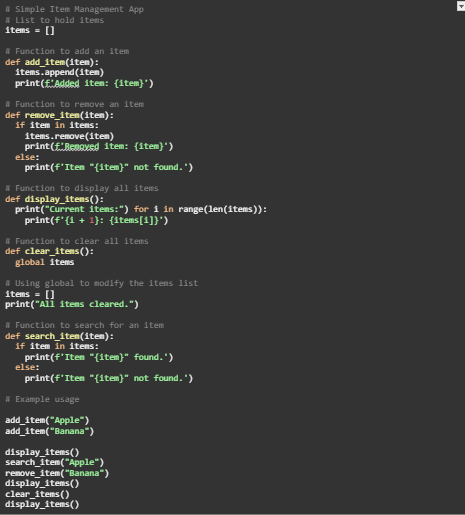}  
        \caption{Contextual Understanding Input Example }
        \label{fig:cpic3}
    \end{minipage}
    \hfill
    \begin{minipage}{0.45\textwidth}
        \centering
        \includegraphics[width=\textwidth]{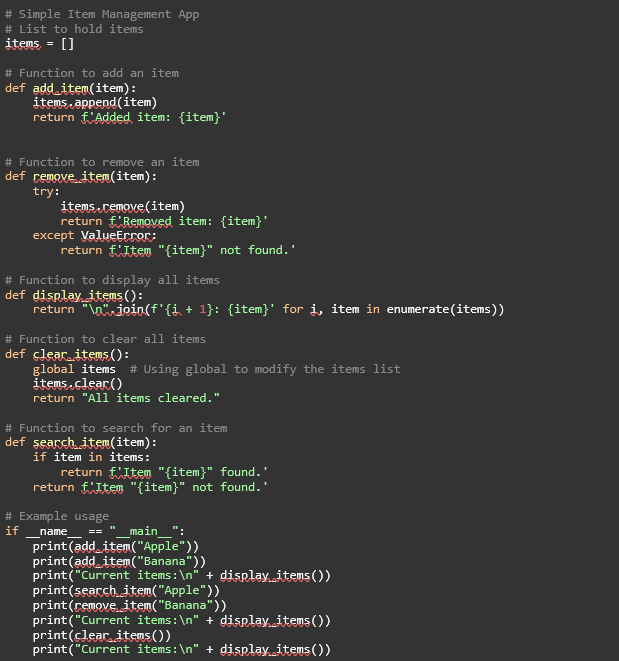}  
        \caption{ChatGpt Reply Contextual Understanding}
        \label{fig:cpic4}
    \end{minipage}
\end{figure*}

\textbf{D. Efficient Multi-Line Edits} \\
ChatGPT can handle formatting, code restructuring, and inserting relevant comments, managing multiple adjustments simultaneously. \\
\textit{Prompt:} "Update all my print statements to use Python's logging module instead." \href{https://chatgpt.com/share/66fa339c-8644-8009-b678-d9c7c6527a72}{Chat Link}

\begin{figure}[h!]
    \centering
    \includegraphics[width=0.6\textwidth, height=5cm]{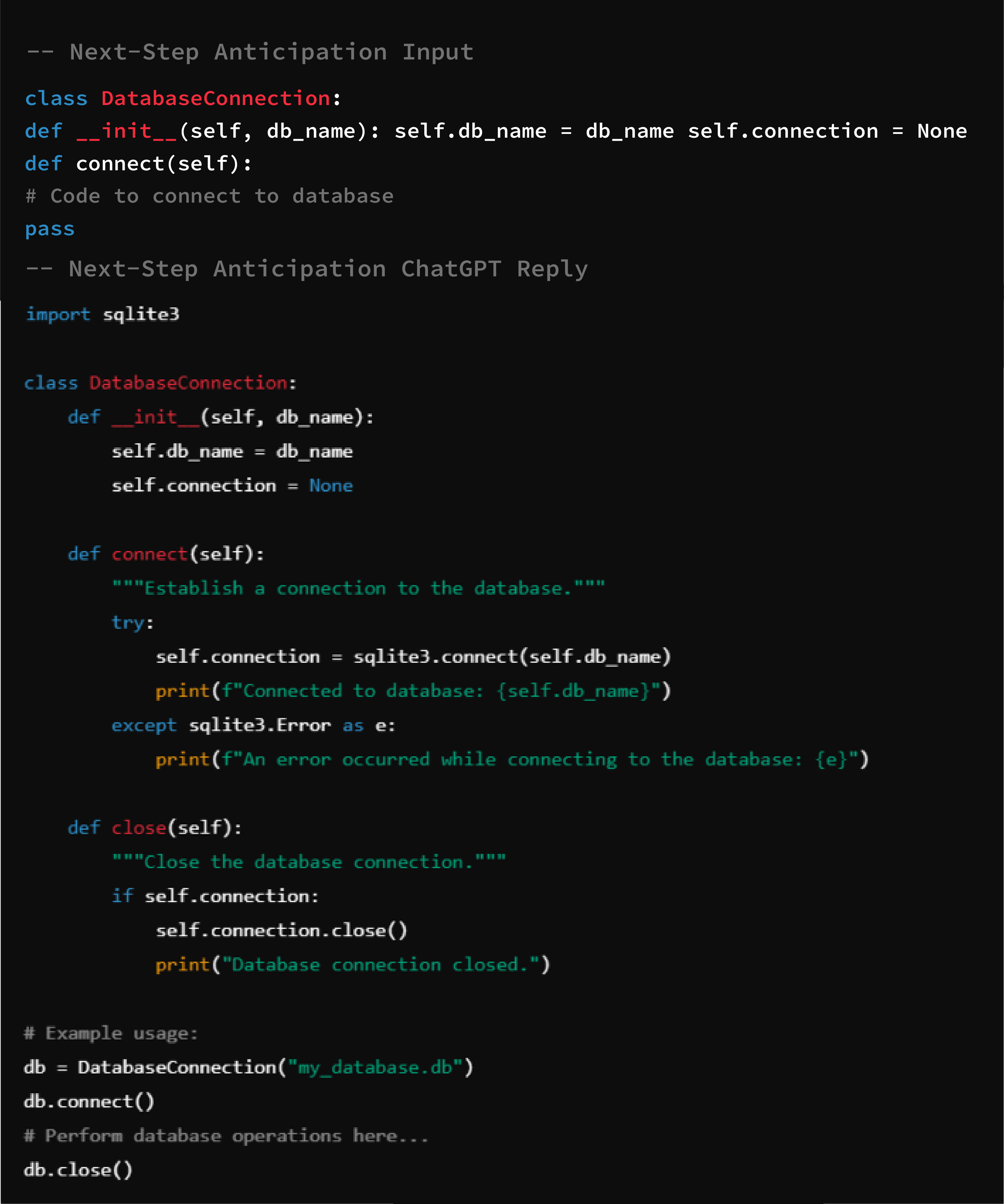}  
    \caption{Next-Step Anticipation Input and ChatGPT Output}
    \label{fig:cpic6}
\end{figure}

\textbf{E. Next-Step Anticipation} \\
ChatGPT often anticipates what you might need next based on the previous commands or coding patterns, such as by suggesting the next logical function, structure, or implementation. \\
\textit{Prompt:} " If write a class for handling database connections, ChatGPT might automatically suggest methods for closing connections, handling exceptions, or creating  queries." \href{https://chatgpt.com/share/66fa3241-b198-8009-b8d8-0cb58f807a0e}{Chat Link}.
\begin{figure}[h!]
    \centering
    \includegraphics[width=0.6\textwidth]{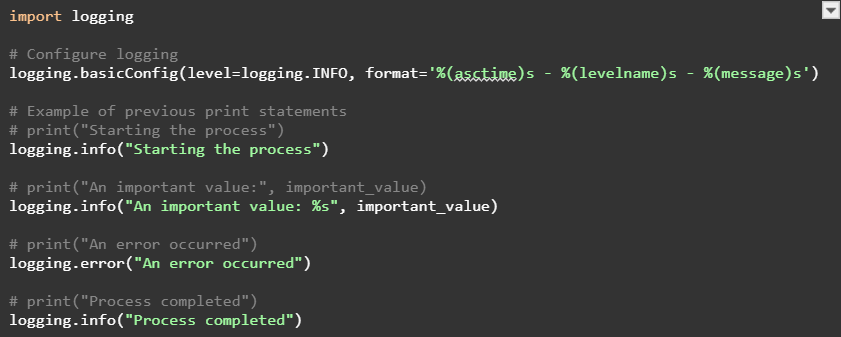}  
    \caption{Adaptive Code Generation Example}
    \label{fig:cpic5}
\end{figure}

\textbf{F. Iterative Debugging and Optimization}

When debugging code, ChatGPT is not merely reactive; it offers proactive suggestions based on detected errors or inefficiencies in code.

\textit{Prompt:} "Why is my code running slowly?".

\textit{ChatGPT Reply:} "After analyzing your code, ChatGPT might suggest optimizations such as reducing the time complexity of a nested loop or replacing a recursive function with an iterative approach to improve performance."

\textbf{G. Enhancing Code Quality and Best Practices}

ChatGPT helps to enforce best practices for writing clean, readable, and efficient code. It emphasizes the importance of well-structured documentation, meaningful variable names, and modular code design to improve overall code quality.

\textit{Prompt:} "How can I improve this function?".

\textit{ChatGPT Reply:} "In addition to refactoring the code for better efficiency, ChatGPT may suggest adding docstrings, meaningful comments, and adopting consistent naming conventions to enhance the function's readability and maintainability.

\textbf{H. Third-Party API Integration}
ChatGPT aids in seamlessly integrating third-party APIs into projects, providing examples of how to authenticate, send requests, and handle responses.\href{https://chatgpt.com/share/66fa3d75-7770-8009-a48d-9d82f6d774ae}{Chat Link}

\textit{Example:} "Integrating a weather API?".

\textit{Prompt:} “How can I fetch weather data from an external API?”.

\begin{figure}[h!]
    \centering
    \includegraphics[width=0.6\textwidth]{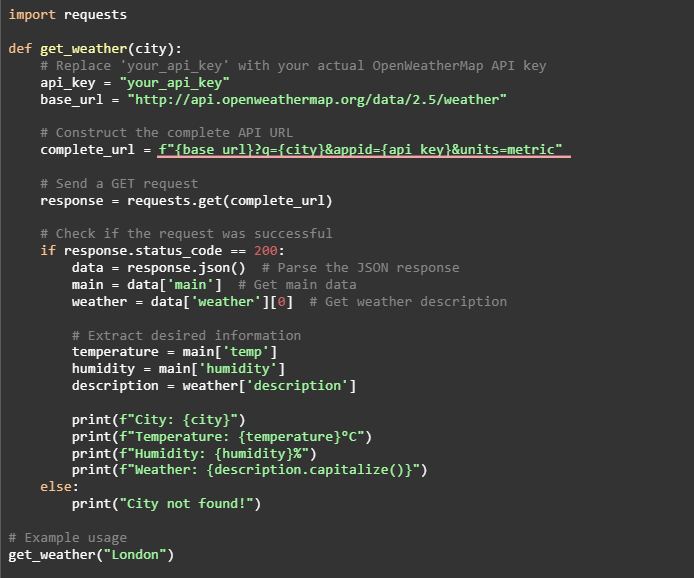}  
    \caption{Third-Party API Integration Example}
    \label{fig:cpic8}
\end{figure}

\textbf{I. Error Correction and Smart Rewrites}
ChatGPT automatically detects and suggests corrections for minor syntax mistakes and typing errors, refining code effectively. It excels at pinpointing errors within code by analyzing the syntax and logic. It can recognize issues that might be overlooked during manual reviews, helping developers maintain code quality \href{https://chatgpt.com/share/66fa36cf-5f30-8009-88d7-30469c69872e}{Chat Link} .

\textit{Example:} "If you provide a function that has a logical error, such as not handling edge cases, ChatGPT can help identify that?".

\textit{Prompt:} “What’s wrong with this code?”.

\begin{figure}[h!]
    \centering
        \includegraphics[width=0.6\textwidth]{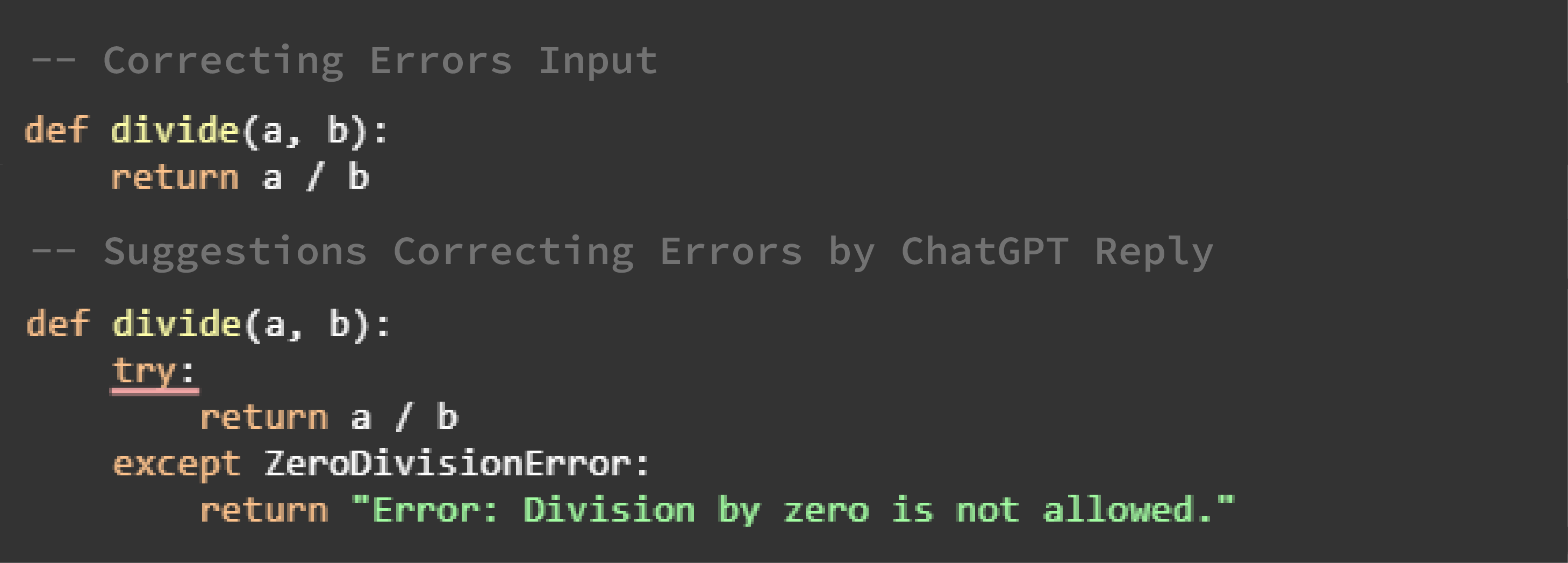}  
        \caption{Correcting Errors Input and Chatgpt Reply}
        \label{fig:cpic10}
    \end{figure}

Once errors are identified, ChatGPT provides specific corrections tailored to the context of your code. This includes syntax corrections, logical fixes, and optimizations.

\textit{Example:}" If your code snippet is meant to calculate the sum of a list but fails, ChatGPT can provide corrections."
\href{https://chatgpt.com/share/66fa3896-033c-8009-8af6-020501dda63e}{Chat Link}

\textbf{J. Suggest alternative methods for Code}
Beyond fixing specific errors, ChatGPT suggests alternative methods and best practices for error handling and code optimization.\href{https://chatgpt.com/share/66fa3927-3b20-8009-ae24-c1832b9af14c}{Chat Link}.

\textit{Example:} "Suggesting a try-except block for error handling?".

\textbf{K. Generate Test Cases for Unit Testing} \\
ChatGPT not only highlights errors but also recommends structured approaches to resolving them. This could involve: \href{https://chatgpt.com/share/66fa3c4a-d7a0-8009-b4be-6d8104c7806d}{Chat Link}
\begin{itemize}
    \item Breaking down complex functions into simpler ones.
    \item Suggesting unit tests to ensure correctness.
\end{itemize}

\textit{Example:} "Suggesting a try-except block for error handling?"

\textbf{L. Code Refactoring and Optimization} \\
ChatGPT can Refactoring the code as well as can Optimize the code. \href{https://chatgpt.com/share/66fa8a01-a2c4-8009-8706-ba4d2ac0e1a5}{Chat Link}

\textit{Example:} "If you provide a nested loop for searching through a list, ChatGPT might suggest using a more efficient algorithm, like binary search or a hash table for faster lookups.?"

\textbf{M. Suggest New Frameworks or Libraries} \\

ChatGPT is a helpful tool for providing code examples in different frameworks and libraries, making coding easier for developers. It can generate code, help debug issues, and explain complex programming concepts. This is useful for both beginners and experienced programmers, as it simplifies coding and helps them learn new technologies faster.\href{https://chatgpt.com/share/66fa8a8a-6328-8009-aa3b-927d60685761}{Chat Link}

\textit{Example:} "If you want to switch a web project from Flask to FastAPI, ChatGPT can translate your code and explain how to use the newer framework.?"

In many organizations, legacy codebases require updates to align with modern frameworks and practices. ChatGPT helps migrate legacy code by suggesting best practices, refactoring strategies, and identifying deprecated functions.

\textbf{N. Converting a legacy Flask app to a newer version} \\

ChatGPT is Updating an old Flask application to a newer version requires careful planning to maintain its functionality and quality. The process can follow best practices from software modernization frameworks, which focus on preserving the application's core value while upgrading the technology stack. \href{https://chatgpt.com/share/66fa3cc7-2ef0-8009-bad1-71b75973b4f6}{Chat Link}.

\textit{Example:} “How can I update this endpoint for Flask 2.0?”

\begin{figure}[h!]
    \centering
        \includegraphics[width=0.6\textwidth]{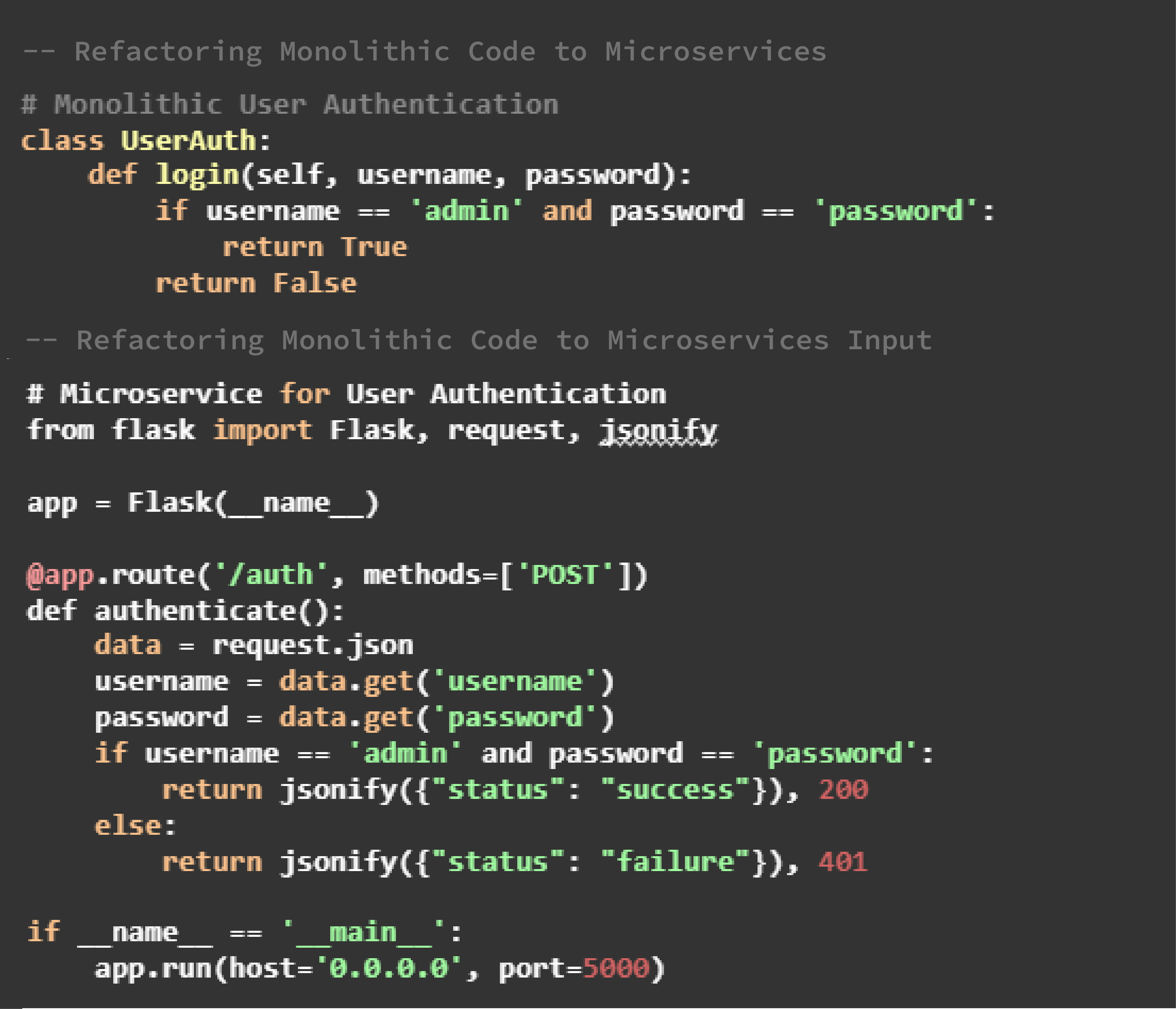}  
        \caption{Converting a legacy Flask app to a newer version Input and Chatgpt Reply}
        \label{fig:cpic13}
    \end{figure}

\textbf{O. Converting an Android RecyclerView to an iOS  UITableView}\\

ChatGPT Converting an Android RecyclerView to an iOS UITableView requires understanding the differences between the two UI frameworks. Using automated tools can make the migration process easier. By applying cross-platform migration technique. \href{https://chatgpt.com/share/66fb76a0-3acc-8009-a920-d5daa3d53f3f}{Chat Link}

\textit{Example:} How do I convert this RecyclerView code into its iOS equivalent?”

\begin{figure}[h!]
    \centering
        \includegraphics[width=0.5\textwidth]{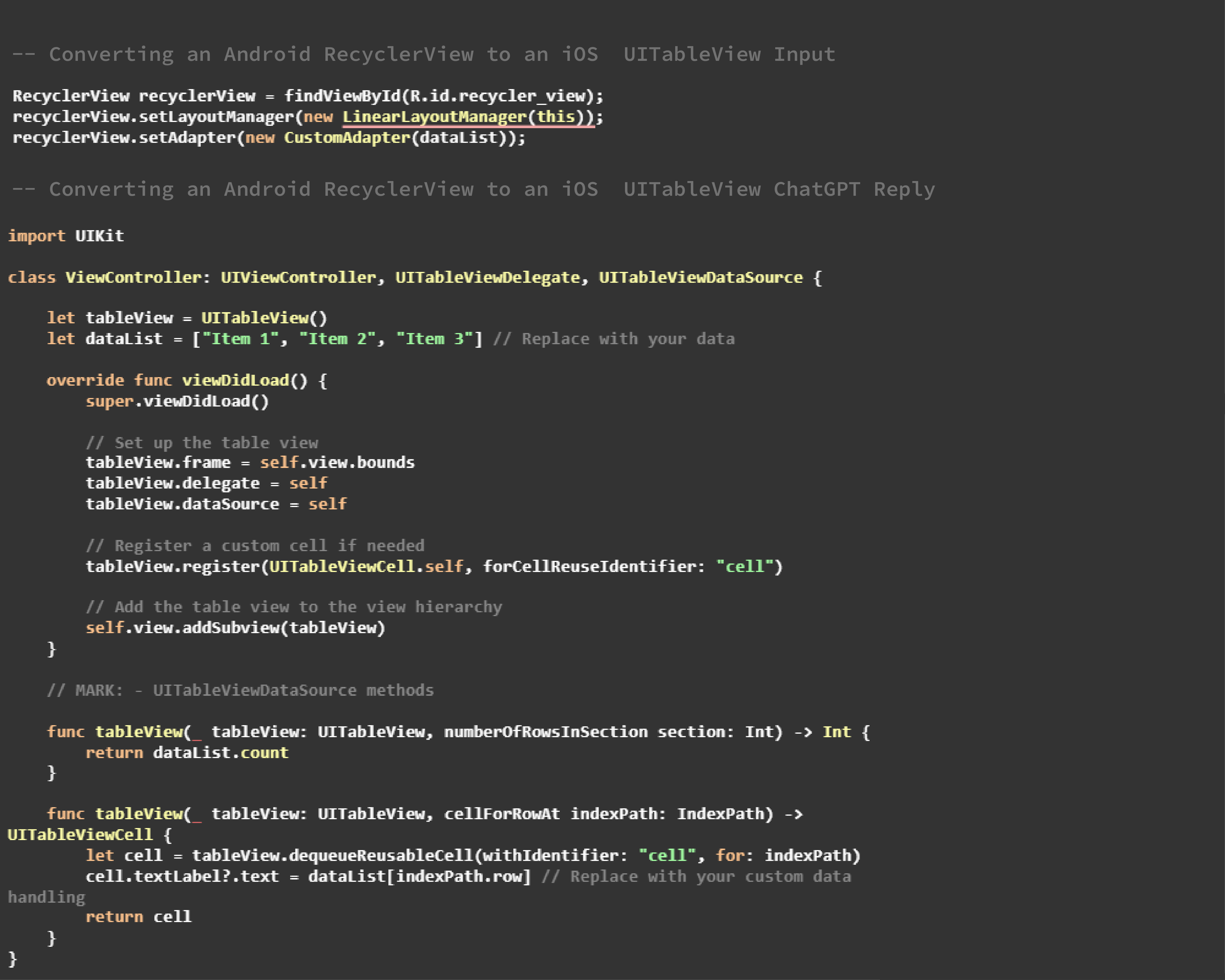}  
        \caption{Converting an Android RecyclerView to an iOS  UITableView Input and Chatgpt Reply}
        \label{fig:cpic14}
    \end{figure}

\textbf{P. Refactoring Monolithic Code to Microservices}\\

ChatGPT can guide you through splitting services, managing API communication, and ensuring scalability.

\textit{Example:}"Extracting a user authentication service from a monolithic system?”
\textit{Prompt:}  “How can I refactor user authentication into a microservice?”

\begin{figure}[h!]
    \centering
        \includegraphics[width=0.5\textwidth]{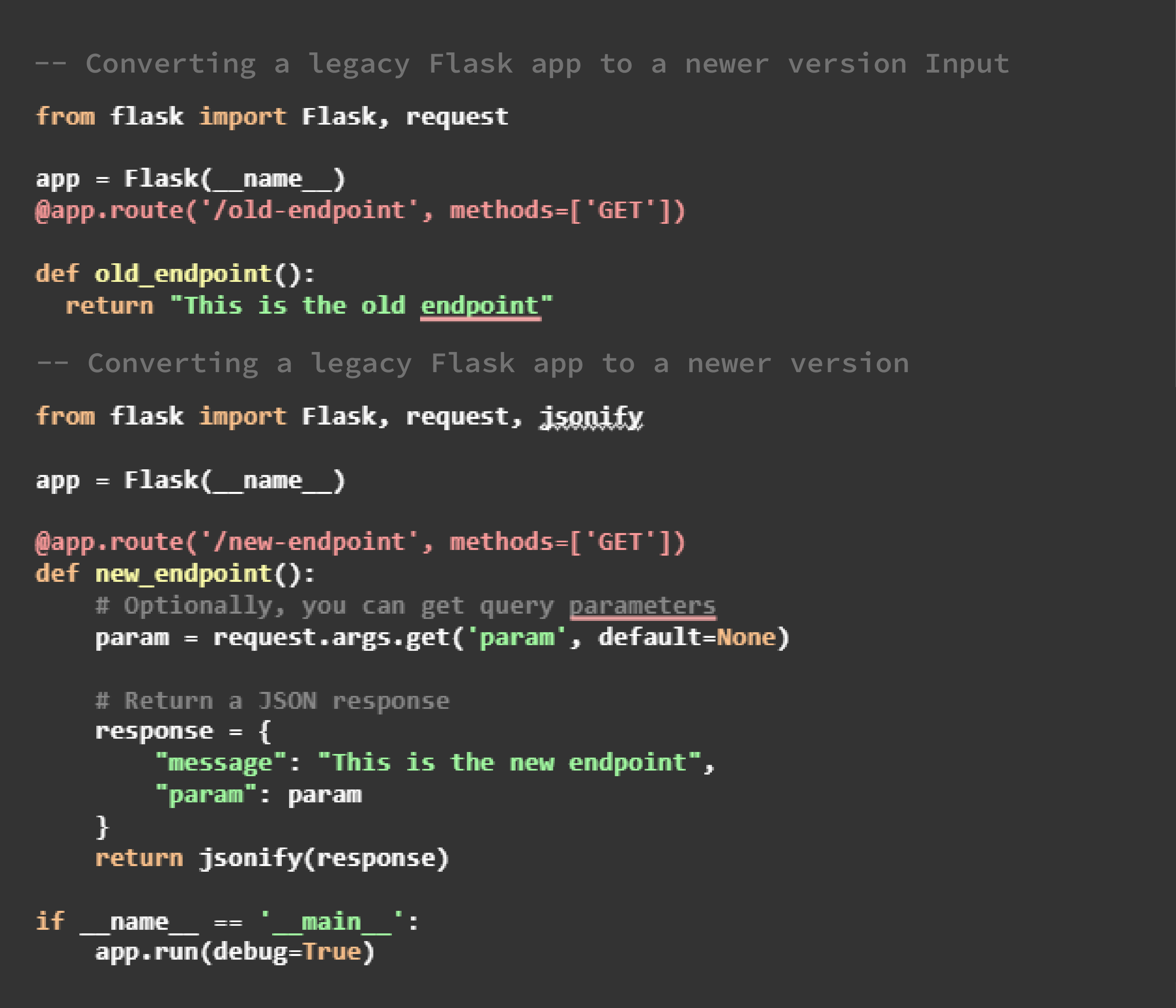}  
        \caption{Refactoring Monolithic Code to Microservices Input and Chatgpt Reply}
        \label{fig:cpic16}
    \end{figure}

\textbf{Q. Database Query Optimization}\\

Optimizing database queries for performance can be challenging. ChatGPT can assist developers in writing optimized SQL queries or suggest indexing strategies.

\textit{Example:} "Optimizing a SQL query for a large dataset?"
\textit{Prompt:} “How can I optimize this query for faster performance?”
\begin{figure}[h!]
    \centering
        \includegraphics[width=0.6\textwidth]{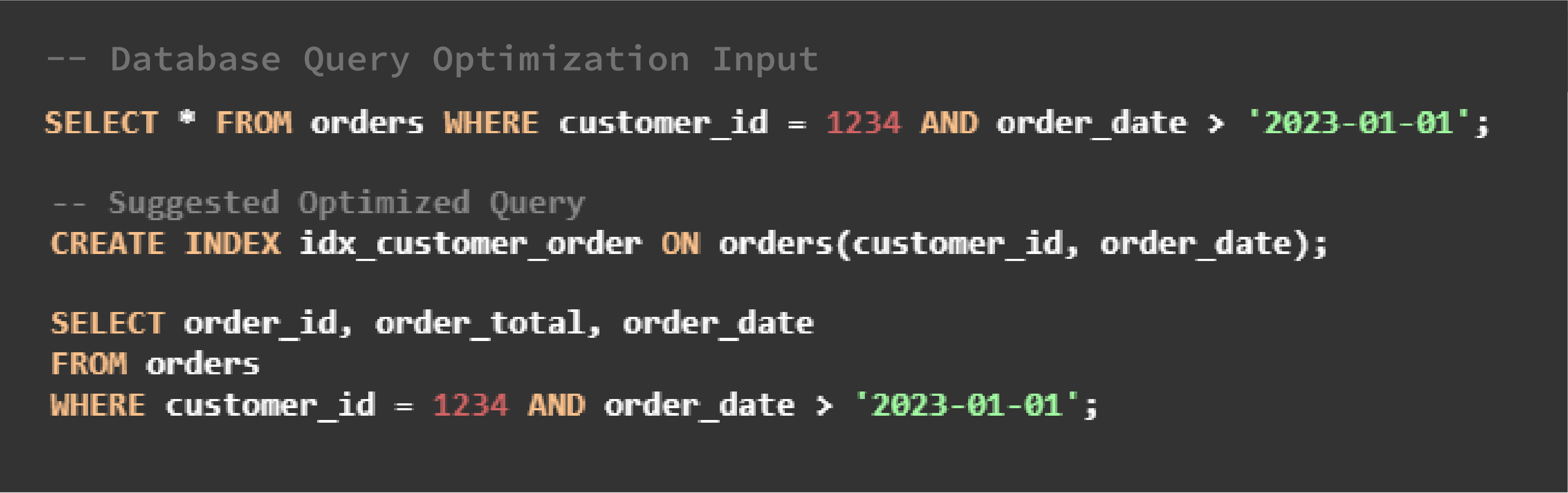}  
        \caption{Database Query Optimization Input and Chatgpt Reply}
        \label{fig:cpic17}
    \end{figure}

\subsection{Cursor AI}

Cursor is an AI code editor that enhances productivity by anticipating edits and providing intelligent coding suggestions. It is a fork of VS Code. This allows us to focus on making the best way to code with AI while offering a familiar code editing experience. It prioritizes privacy with a local storage mode for code and integrates effortlessly with existing tools and workflows.

\subsubsection{Evaluation}

AI-assisted coding tools like \textbf{Cursor AI} have revolutionized programming by enabling natural language-driven code generation, debugging, and optimization. However, they pose risks, particularly in terms of security vulnerabilities. Since they are trained on vast datasets, including potentially insecure or outdated code, they can generate vulnerabilities like SQL injection, XSS, weak authentication methods, and poor error handling \cite{anala2024dangers}. Developers must follow security best practices to avoid deploying unsafe code. Users have reported challenges with Cursor AI when using multiple files \cite{cursor2024privacy}:

\begin{itemize} \item \textbf{File Version Hallucinations:} AI may mistake current file versions for outdated ones, leading to redundant or incorrect suggestions \cite{cursorforum2024hallucinations}. \item \textbf{Contextual Gaps:} Without full folder analysis, AI struggles to understand multi-file contexts, leading to irrelevant or repetitive suggestions \cite{cursorforum2024hallucinations}. \end{itemize}
These issues highlight inefficiencies in the tool's management of file context, affecting usability and performance. Additionally, misuse of Cursor AI can degrade code quality. While the generated code may appear correct initially, small errors and inefficiencies can accumulate, undermining project quality. Cursor AI also has limitations, particularly with complex projects \cite{cursorai2024dangers}. While it handles simple tasks well, it struggles with larger project structures, leading to incorrect suggestions and subtle bugs. Although effective for basic code, it delays with advanced needs, such as specific business logic or custom frameworks \cite{cursorai2024dangers}, creating a false sense of security for new developers who may rely too heavily on AI-generated code. Another concern is data privacy, especially for projects involving sensitive information. Despite Privacy Mode, users have raised questions \cite{cursor2024privacy} about whether the tool still stores data, potentially violating NDAs. Enabling "Privacy Mode" ensures that no code is stored, except for temporary prompt data retained by OpenAI and Anthropic for 30 days \cite{cursor_privacy2}.
 
\subsubsection{User Feedback Overview}
A user survey indicated that 15.2\% of users rated its generated code quality at 50\%, with 9.1\% rating it between 60\% and 80\%. Cursor AI’s error identification capabilities were found to effectively reduce debugging time. Additionally, 69.7\% of users reported enhanced collaboration, though 18.2\% found it ineffective for collaborative tasks. Reliability and performance ratings showed that 48.5\% of users gave it a 4, while 33.3\% rated it a 3.

\subsubsection{Security Analysis}
Cursor AI uses subprocessors (e.g., AWS, Fireworks, OpenAI, Anthropic, Google Cloud Vertex API) and cloud services to deliver its AI features\cite{cursor_security}. When we disable privacy mode, Cursor AI gathers telemetry and usage data, such as code snippets and editor actions, to enhance its AI capabilities\cite{cursor_privacy2}. Cursor AI temporarily caches and encrypts this data on servers, but neither permanently stores nor uses it for training purposes. However, if privacy mode is enabled, no code data is stored or retained by Cursor or any third party\cite{cursor_privacy2}. When using an API key, requests pass through Cursor's backend for the final prompt construction. When enabling code indexing, Cursor uploads small portions of code for embedding calculations and deletes the raw code after completing the process. Cursor stores only the embeddings and associated metadata, including file names and hashes \cite{cursorforum2024hallucinations}. Furthermore, file contents are temporarily cached on servers, encrypted with unique client keys, and are not utilized for training when Privacy Mode is enabled\cite{cursorai2024dangers}.

\subsubsection{Case study}
\textbf{A. Code Generation} 

Cursor predicts your next steps based on recent changes, tracks the codebase, and suggests relevant code, enhancing development efficiency. It adapts to past interactions, improving the accuracy of its suggestions and ensuring context-aware edits.

\begin{figure}[h!]
    \centering
    \includegraphics[width=0.6\textwidth]{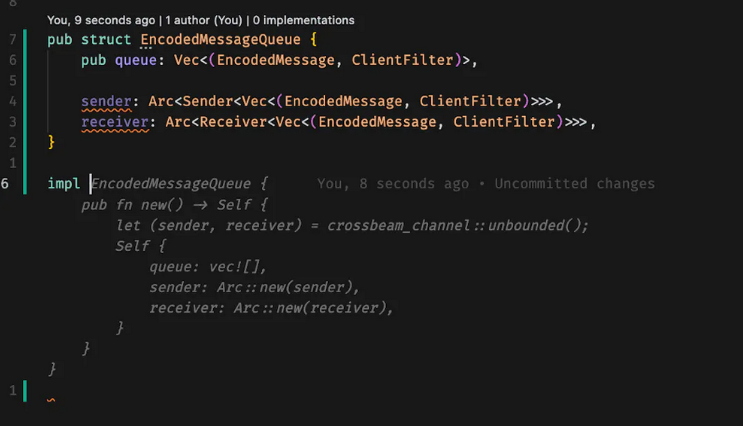}  
    \caption{Code Generation by cursor }
    \label{fig:crpic1}
\end{figure}

\textbf{B. Multi-Line Edits} 
Cursor suggests multiple edits at once, saving time and reducing errors. It intelligently rewrites code blocks for better readability and performance, streamlining the development process.

\begin{figure}[h!]
    \centering
    \includegraphics[width=0.6\textwidth]{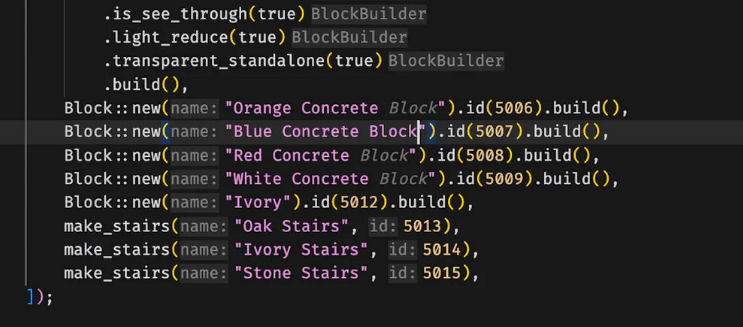}  
    \caption{Multi-Line Edits by cursor }
    \label{fig:crpic2}
\end{figure}

\textbf{C. Smart Rewrites} 
Cursor fixes errors in your code, improving readability and ensuring consistency in coding style. It refactors your code for better performance, helping to prevent potential bugs and save time on corrections.

\begin{figure}[h!]
    \centering
    \includegraphics[width=0.6\textwidth]{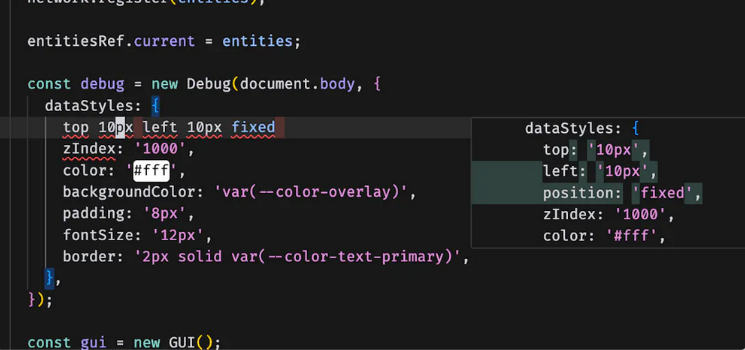}  
    \caption{Smart Rewrites by cursor }
    \label{fig:crpic3}
\end{figure}

\textbf{D. Cursor Prediction} By predicting your next cursor position, Cursor enhances navigation and streamlines coding. It anticipates movements, making it easier to navigate a large codebase and improving workflow efficiency.

\begin{figure}[h!]
    \centering
    \includegraphics[width=0.6\textwidth]{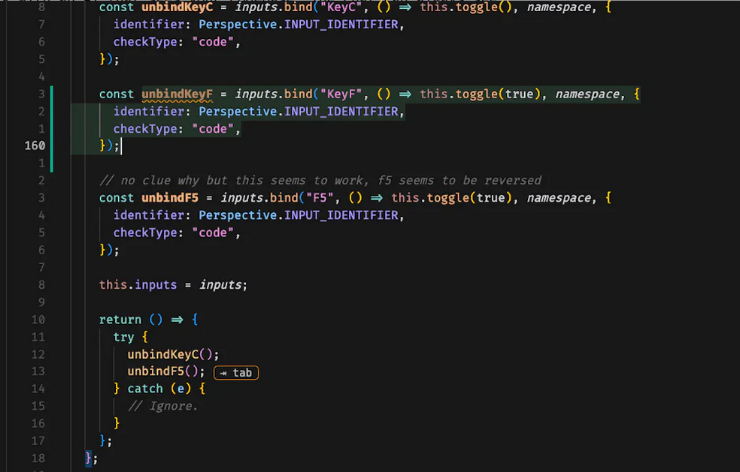}  
    \caption{Cursor Prediction}
    \label{fig:crpic4}
\end{figure}

\textbf{E. Identifying Errors and Fix Suggestions}

Cursor AI can identify errors effectively. In the compile time, it gives the suggestion by AI which is an important feature in cursor AI.

\begin{figure}[h!]
    \centering
    \includegraphics[width=0.6\textwidth]{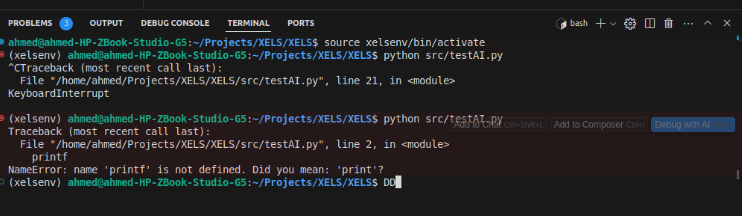}  
    \caption{Identifying Errors and Fix Suggestions}
    \label{fig:crpic5}
\end{figure}

Cursor Composer is an advanced AI feature in the Cursor editor that simplifies multi-file editing and full application development. It allows developers to provide high-level instructions for building or modifying entire applications while accounting for the project structure. Its key features include multi-file editing, app generation, contextual understanding, and interactive refinement. Cursor integrates custom API keys for enhanced flexibility and provides effective error resolution with interactive fixes, ensuring maintainable and functional code.

\subsection{Codeium AI}

Codeium AI is a cutting-edge AI-driven tool that revolutionizes software development. It improves code quality, automates testing processes, and integrates effortlessly with popular coding platforms such as VSCode and JetBrains IDEs.

\subsubsection{Evaluation}
Some companies avoid AI tools because of security concerns, and developers with unique workflows may have trouble using these assistants. Codeium solves these problems by allowing on-premise deployment, making it secure and customizable for specific projects. This ensures that developers can maintain control over their code while using the tool. Codeium is also fully integrated and works with other development tools, so it can be used alongside existing workflows without risk. By offering local deployment, Codeium ensures that sensitive data stays within the company, reducing the chances of outside security breaches \cite{wu2024unveiling}\cite{oh2024poisoned}. This feature also helps companies meet legal requirements for data privacy and security \cite{codeium_security}. Additionally, the tool can be customized to fit specific coding practices, which ensures the AI works according to company standards. With over 300,000 free users and 100 enterprise clients, Codeium is widely trusted for its combination of productivity and security 
\cite{agarwal2024codemirage}\cite{alawida2024unveiling}\cite{privacydesigner2024gdpr}. It enables developers to work faster while ensuring their data is safe.

\subsubsection{User Feedback Overview}
Surveys revealed that Codeium significantly streamlined workflows, with time savings highlighted as a major benefit. Its generated code quality was rated with peaks at 40\% and 60\%, each at 12.5\%, while higher ratings above 75\% were relatively scarce. Debugging time reduction was another strength, as users found it effective for resolving complex issues. Collaboration and support metrics indicated that 57.1\% of users experienced improved collaboration, though 32.1\% reported it as ineffective for teamwork. Reliability and performance ratings showed that 58.1\% of users rated it a 3, with 16.1\% and 6.5\% giving it ratings of 4 and 5, respectively.

\subsubsection{Security Analysis}

Codeium AI \cite{privacydesigner2024gdpr} processes various types of data, including code snippets, metadata, user authentication details, and model configurations, which are transmitted through its system using cloud infrastructure. This raises concerns about data security and potential leaks, such as the risk of personal data embedded in code, cross-border data transfers, unauthorized access to source code, AI-generated vulnerabilities, and intellectual property violations. Codeium mitigates these risks by implementing strict code review guidelines, employing contractual safeguards for cross-border transfers, using advanced user authentication and access control, conducting vulnerability scanning, and ensuring proper license verification. To prevent data leaks, Codeium \cite{codeium_security} emphasizes encryption, offers self-hosted deployment options for enterprise users, and restricts data access to authorized personnel. Through these measures, Codeium aims to ensure responsible data handling and privacy protection, urging organizations to implement continuous monitoring and internal controls.

\subsubsection{Case study}

\textbf{A. Code Generation}
Codeium's code generation feature enables code generation simply by describing tasks in natural language. Using natural language processing, it creates high-quality code that matches with needs.  It works for various programming languages, such as Go, HTML, or Unity.  This feature makes coding faster and more accessible, even for complex tasks. A best-in-class proprietary model, trained from scratch to optimize for speed and accuracy, powers this feature.

\begin{figure}[h!]
    \centering
    \includegraphics[width=0.6\textwidth]{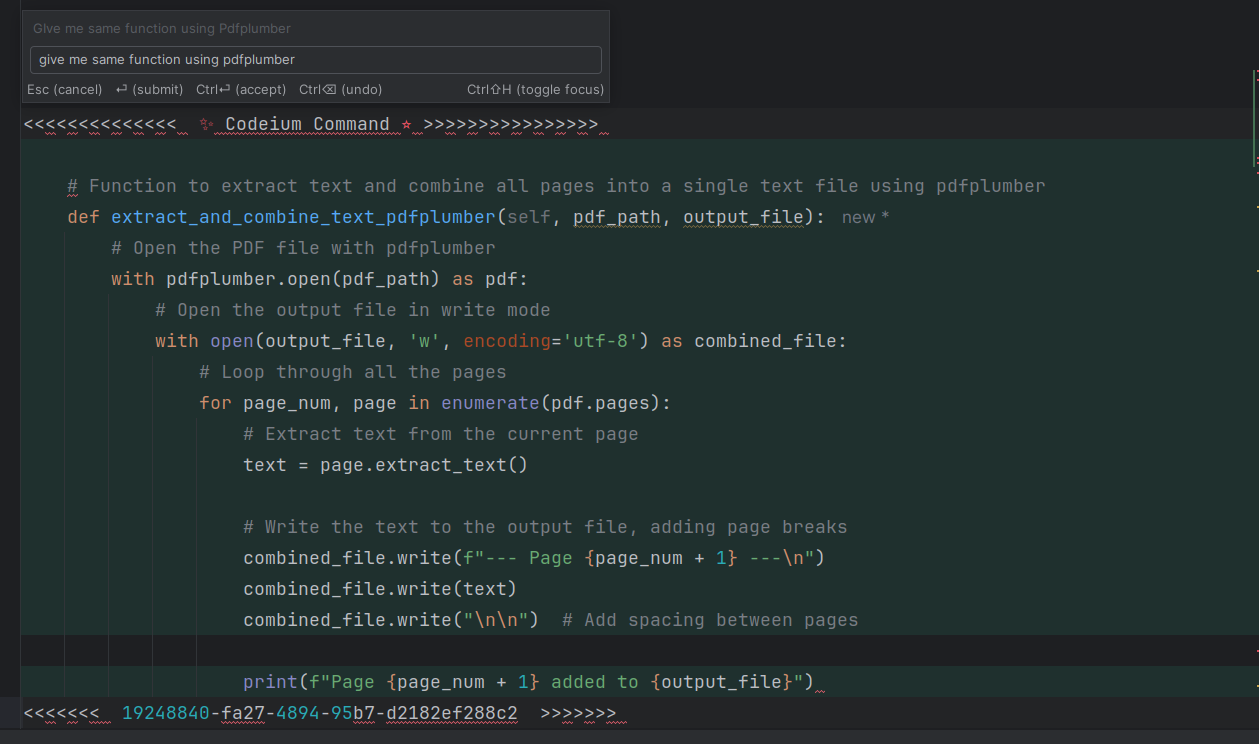}  
    \caption{CodeiumAI Code Generation }
    \label{fig:codiumcodegen}
\end{figure}

\textbf{B. Code Refactoring}
It has strong code refactoring options. There are multiple options and features for code refactoring, which are shown below.

\begin{figure}[h!]
    \centering
    \includegraphics[width=0.6\textwidth]{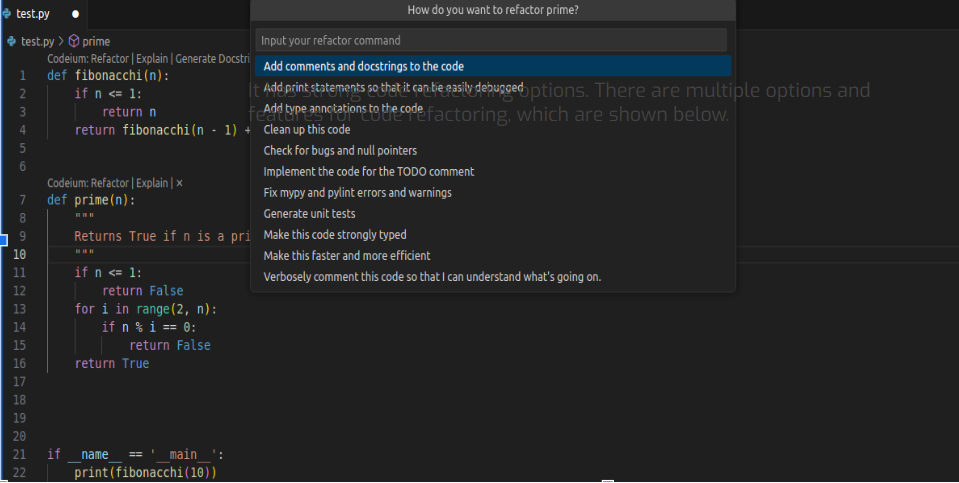}  
    \caption{CodeiumAI Code Refactoring }
    \label{fig:codiumrefactor}
\end{figure}

\textbf{C. Code Debugging} 
Codeium offers real-time debugging through its integrated chat support. It leverages advanced AI models for precise code analysis and bug detection. Supports debugging across 70+ programming languages, ensuring compatibility with diverse projects. Enhances debugging with smart autocomplete suggestions tailored to resolving issues more efficiently.

\begin{figure}[h!]
    \centering
    \includegraphics[width=0.7\textwidth]{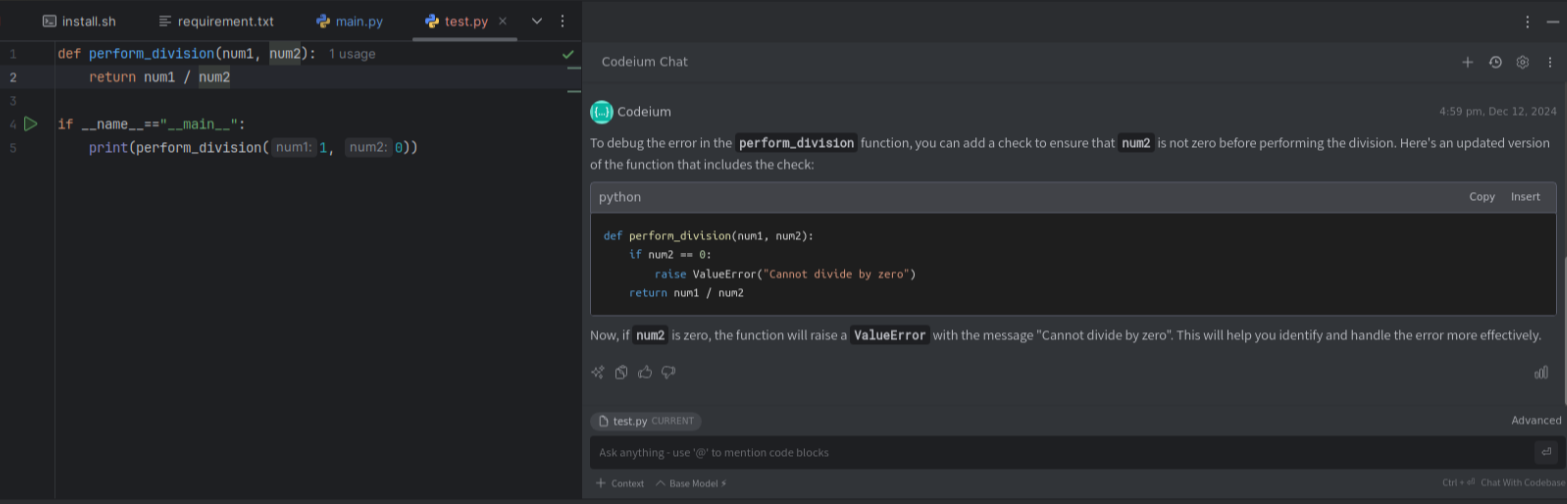}  
    \caption{CodeiumAI Code Debugging }
    \label{fig:codiumdebug}
\end{figure}
\textbf{D. Code explanation} 
Codeium's code explanation feature helps users understand code easily by providing clear, concise explanations. It uses natural language generation and code analysis to explain the purpose, logic, and meaning of code snippets or expressions. Users can input code in various languages like Java, R, Python, or C\# to get instant explanations. Codeium supports multiple programming languages and editors. To use the feature, press \textbf{Ctrl+Shift+Space} or click the Codeium icon in your editor. A window will appear where you can paste or select code, and Codeium will generate an explanation. This feature makes coding easier.

\begin{figure}[h!]
    \centering
    \includegraphics[width=0.6\textwidth]{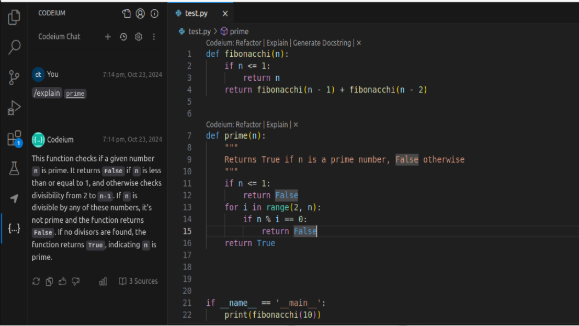}  
    \caption{CodeiumAI Code explanation }
    \label{fig:codiumcodeexplain}
\end{figure}

\textbf{E. Code Completion}

Codeium's code completion feature helps users write code faster by suggesting relevant keywords, functions, and parameters. It uses deep learning to provide context-aware suggestions based on code. For example, in Python, it suggests correct syntax and arguments; in SQL, it recommends tables and columns; in Excel, it offers suitable functions and formats. This feature supports multiple programming languages and editors. 

\begin{figure}[h!]
    \centering
    \includegraphics[width=0.6\textwidth]{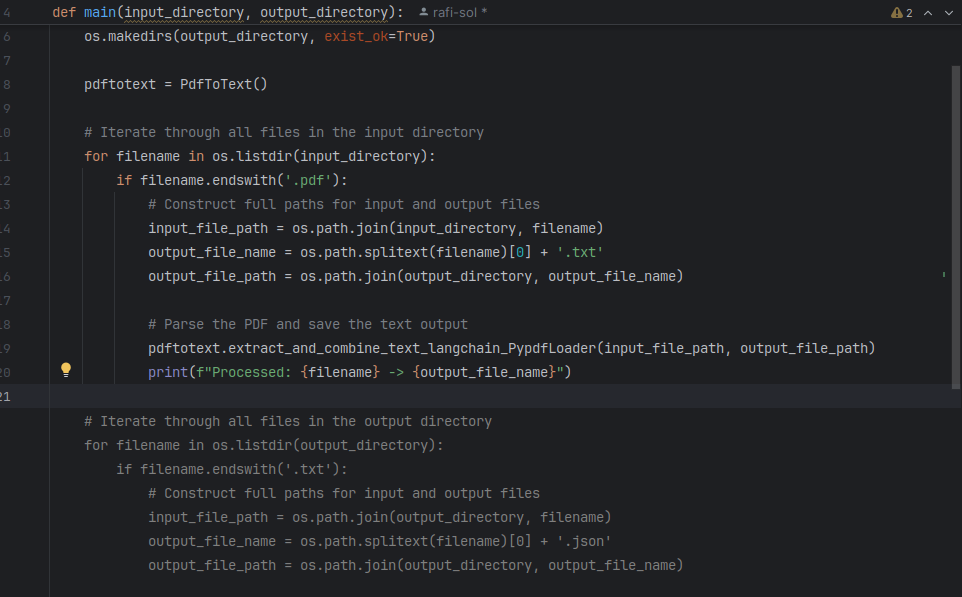}  
    \caption{CodeiumAI Code explanation }
    \label{fig:codiumcodeautocomplete}
\end{figure}

\textbf{F. Supercomplete Features}
In October 2024, Codeium introduced a new feature called Supercomplete. Supercomplete is a passive AI that shows the changes insertions, deletions, and edits that match your next action in a pop-up next to the text in your editor. It works independently of where your cursor is positioned.